\documentclass[preprint,showpacs,preprintnumbers,amsmath,amssymb]{revtex4}

\usepackage{graphicx}
\usepackage{dcolumn}
\usepackage{amsmath}
\usepackage{amssymb}

\newcommand{\bvec}[1]{\mbox{\boldmath $#1$}}
\newcommand{\D}{\delta }

\def\vq{{\bf q}}

\def\vk{{\bf k}}
\def\vQ{{\bf Q}}

\def\vr{{\bf r}}
\def\vS{{\bf S}}

\def\bra{\langle}
\def\ket{\rangle}

\newcommand{\eq}[1]{Eq.~(\ref{#1})}
\newcommand{\fig}[1]{Fig.~\ref{#1}}
\newcommand{\be}{\begin{equation}}
\newcommand{\ee}{\end{equation}}
\newcommand{\bea}{\begin{eqnarray}}
\newcommand{\no}{\nonumber}
\newcommand{\eea}{\end{eqnarray}}
\newcommand{\bean}{\begin{eqnarray*}}
\newcommand{\eean}{\end{eqnarray*}}
\newcommand{\bfi}{\begin{figure}}
\newcommand{\efi}{\end{figure}}
\newcommand{\bc}{\begin{center}}
\newcommand{\ec}{\end{center}}
\newcommand{\ba}{\begin{array}}
\newcommand{\ea}{\end{array}}

\begin{document}


\title{Magnetic excitations and their anisotropy in YBCO: \\ 
slave-boson mean-field analysis of bilayer $t$-$J$ model} 

\author{Hiroyuki Yamase and Walter Metzner} 
\affiliation{Max-Planck-Institute for Solid State Research, 
Heisenbergstrasse 1, D-70569 Stuttgart, Germany}


\date{\today}

\begin{abstract} 
We perform a comprehensive analysis of the dynamical magnetic 
susceptibility $\chi(\vq,\,\omega)$ in the slave-boson mean-field 
scheme of the bilayer $t$-$J$ model.  
We use model parameters appropriate for YBa$_{2}$Cu$_{3}$O$_{6+x}$ 
(YBCO), a typical bilayer high-$T_{c}$ cuprate compound well 
studied by neutron scattering experiments. 
In the $d$-wave pairing state, the strongest magnetic spectral weight 
appears at $\vq=\vQ\equiv (\pi,\,\pi)$ and $\omega=\omega_{\vQ}^{\rm res}$, 
and spreads into a diamond-shaped shell around $\vQ$ in $\vq$ space 
for $\omega<\omega_{\vQ}^{\rm res}$. 
This weight is due to a collective  mode, namely a particle-hole 
bound state, which has a downward $\omega$ versus $\vq$ dispersion
around $\vQ$. 
Within the high intensity shell, the incommensurate (IC) signals at 
$\vq=(\pi,\,\pi\pm 2\pi\eta)$ and $(\pi\pm 2\pi\eta,\,\pi)$ tend to 
be stronger than the diagonal incommensurate (DIC) signals at 
$\vq=(\pi\pm 2\pi\eta',\,\pi\pm 2\pi\eta')$, 
especially for a large hole density $\D$. 
For $\omega \ll \omega_{\vQ}^{\rm res}$ the IC signals completely disappear 
and the weight remains only around the DIC positions. 
For $\omega > \omega_{\vQ}^{\rm res}$ strong signals of Im$\chi(\vq,\,\omega)$ 
tracing an upward dispersion are found and interpreted as an overdamped 
collective mode near $\omega_{\vQ}^{\rm res}$.  
In the normal state, Im$\chi(\vq,\,\omega)$ has a broad peak at $\vq=\vQ$. 
That is, the IC and DIC signals appear only in the $d$-wave pairing state. 
We also study effects of a small orthorhombic anisotropy, which is intrinsic 
in untwinned YBCO crystals. 
Because of electron-electron correlations favoring $d$-wave shaped 
Fermi surface deformations ($d$FSD), 
we expect an enhanced anisotropy of magnetic excitation spectra. 
This effect is particularly pronounced for low $\D$ and at relatively high 
temperature. 
The present theory provides a rather detailed microscopic explanation of
the most salient properties of magnetic excitations observed in YBCO. 
\end{abstract}

\pacs{74.25.Ha, 74.72.Bk, 74.20.Mn, 71.10.Fd}
\maketitle

\section{Introduction} 
The undoped parent compounds of high-$T_{c}$ cuprate superconductors 
are antiferromagnetic Mott insulators. 
With carrier doping the antiferromagnetism disappears and a superconducting 
state with a high transition temperature is realized.  
Antiferromagnetic (AF) correlations however remain even 
in the superconducting state and the understanding of magnetic properties 
has been widely recognized as a major issue in the theory of high-$T_{c}$ 
cuprates. 

Magnetic correlations are directly measured by inelastic neutron 
scattering experiments. Most of the experiments were performed for the
cuprate superconductors YBa$_{2}$Cu$_{3}$O$_{y}$ (YBCO) and 
La$_{2-x}$Sr$_{x}$CuO$_{4}$ (LSCO). 
Early studies suggested that YBCO had commensurate antiferromagnetic 
correlations, that is, the peak signal in the imaginary part of the dynamical
magnetic susceptibility $\chi(\vq,\omega)$ at various frequencies appeared 
at $\vq=\vQ\equiv (\pi,\,\pi)$,\cite{rossatmignod91} while LSCO had 
incommensurate (IC) correlations in the sense that the peak shifted to 
$\vq=(\pi\pm2\pi\eta,\,\pi)$ and $(\pi,\,\pi\pm2\pi\eta)$, 
where $\eta$ parametrizes the degree of
incommensurability.\cite{thurston89,cheong91,mason92,thurston92} 
In 1998, measurements with improved resolution revealed that the seemingly 
commensurate signal of YBCO was actually composed of four peaks at 
incommensurate wavevectors.\cite{dai98} 
Indications for an incommensurate signal in YBCO appeared already in several 
earlier observations,\cite{tranquada92,sternlieb94} but remained ambiguous
due to a limited resolution. 
The peak positions were identified as $\vq=(\pi\pm2\pi\eta,\,\pi)$ and 
$(\pi,\,\pi\pm2\pi\eta)$,\cite{mook98,arai99} the same geometry as in LSCO. 
The IC peaks, however, appeared only below the superconducting transition 
temperature $T_{c}$ or possibly below the pseudogap 
temperature $T^{*}$,\cite{dai98,arai99,bourges00,dai01,stock04} 
in sharp contrast to LSCO, where the IC signals were observed 
at temperatures well above $T_{c}$.\cite{thurston89,aeppli97}

The size of the incommensurability $\eta$ depends on the excitation 
energy $\omega$. 
For YBCO, $\eta$ decreases with increasing $\omega$ and vanishes
at a specific frequency $\omega_{\vQ}^{\rm res}$.
The commensurate, so-called ``resonance peak'' at $\omega_{\vQ}^{\rm res}$ 
was extensively investigated for optimally 
doped\cite{rossatmignod91,mook93,bourges95,fong9596} 
and underdoped samples\cite{bourges95,bourges96,dai96,fong97} before 
the observation of the IC signals in YBCO. 
The resonance peak is now regarded to be continuously connected with 
the IC signals observed at lower frequencies, that is the peak disperses 
smoothly downwards to lower frequencies when $\vq$ is shifted away from 
$\vQ$.\cite{arai99,bourges00,reznik04,pailhes04} 
Above the resonance energy, possible IC structures were 
observed.\cite{arai99} It was recently found that the strongest 
weight appeared at diagonal incommensurate (DIC) positions 
$\vq=(\pi\pm2\pi\eta,\,\pi\pm2\pi\eta)$.\cite{hayden04}
On the other hand, for LSCO the energy dependence of $\eta$ was 
relatively weak compared to YBCO\cite{mason92} 
and such a robustness of IC structures was often contrasted with the
behavior in YBCO. 
However, recent high energy neutron scattering data for 
La$_{2-x}$Ba$_{x}$CuO$_{4}$ with $x=0.125$ revealed that 
the IC peaks dispersed with $\omega$ and merged into a 
commensurate peak around 55meV.\cite{tranquada04} 
Above this energy the data showed weak DIC peaks as in YBCO.  

Early theoretical work\cite{si93,tanamoto93,tanamoto94,liu95} 
pointed out that the differences of magnetic excitations between 
YBCO and LSCO could be traced back to the difference of the Fermi 
surface (FS) shape, which we refer to as the fermiology scenario.    
The FS difference was indeed predicted by LDA band 
calculations\cite{xu87,yu87,park88} and angle-resolved photoemission 
spectroscopy (ARPES) supported the LDA prediction for 
YBCO;\cite{campuzano90,liu92} 
ARPES data for LSCO were not available for a long time. 
After the experimental observation of IC peaks in YBCO, further detailed 
calculations\cite{brinckmann99,kao00,norman00,tchernyshyov01,brinckmann01,
li02,schnyder04} within the same theoretical framework as in the early
days showed that essential features of magnetic excitations in YBCO 
were captured in the fermiology scenario. 

Fermiology theories for LSCO systems were running into problems when 
ARPES data, first reported in 1999,\cite{ino9902} suggested that the 
FS of LSCO looks similar to that of YBCO in the doping region below 
$x \sim 0.15 \,$.\cite{zhou01} 
If so, fermiology scenarios predict that magnetic excitations should be 
essentially the same in both YBCO and LSCO, in contradiction with the 
experimental data. 
This problem was considered within the slave-boson mean-field scheme 
of the $t$-$J$ model.\cite{fukuyama98} 
It was shown\cite{yamase00} that the model has tendencies toward 
orientational symmetry breaking of the square lattice symmetry, 
leading to a $d$-wave shaped FS deformation ($d$FSD): 
the FS expands along the $k_{x}$ axis and shrinks along the $k_{y}$ 
axis or vice versa. 
Assuming a coupling to the low temperature tetragonal lattice 
distortions or their fluctuations, 
one could understand both the 
observed FS shape and the structure of magnetic excitations in LSCO 
consistently.\cite{yamase00,yamase01,yamase02,yamase03} 

An essentially different picture for magnetic excitations was 
proposed in 1995,\cite{tranquada95} the spin-charge stripe scenario, 
according to which IC magnetic excitations are mainly controlled by 
charge-stripe order tendencies in the CuO$_{2}$ plane. 
This scenario is based on the observation of a charge order signal  
whose wavevector is just twice as large as the IC magnetic wavevector. 
Although charge order was observed only in a few high-$T_{c}$ cuprate 
materials with specific 
doping rates,\cite{tranquada95,niemoller99,ichikawa00,fujita02b} and 
in addition the signal was rather weak, the spin-charge stripe scenario 
attracted much interest.\cite{kivelson03} 

The spin-charge stripe scenario predicts one-dimensional 
magnetic signals, that is, IC magnetic peaks should appear either at 
$\vq=(\pi\pm2\pi\eta,\,\pi)$ or $(\pi,\,\pi\pm2\pi\eta)$. 
Such a one-dimensional pattern was actually inferred from experiments 
for partially untwinned YBCO crystals\cite{mook00} 
as a strong support for the stripes. 
However, subsequent neutron scattering studies for almost fully 
untwinned YBCO\cite{hinkov04,hinkov06} revealed that magnetic excitations 
were two-dimensional and had four IC signals. 
The IC pattern however lost fourfold symmetry around $\vq=\vQ$ and 
exhibited some anisotropy between the $q_{x}$ and $q_{y}$ direction. 
There are several possible explanations for this anisotropy. 
(i) The bare band structure effect, due to the orthorhombicity of the 
crystal structure\cite{eremin05,schnyder06} or the CuO chains in 
YBCO.\cite{zhou0405} 
(ii) The tendency to a $d$FSD due to electron-electron correlations,  
which enhances the bare anisotropy.\cite{yamase00,kao05} 
(iii) Effects of charge stripe fluctuations.\cite{vojta06} 
The putative charge stripes should be strongly fluctuating to be
consistent with the experimental observation that the magnetic 
signals form a two-dimensional anisotropic geometry. 

We focus on the possibility of an enhanced anisotropy due to
correlations favoring a $d$FSD.
The $d$FSD tendency is generated by forward scattering interactions 
of electrons close to the FS near $(\pi,\,0)$ and $(0,\,\pi)$. 
It was first found in the $t$-$J$ model\cite{yamase00} and Hubbard 
model,\cite{metzner00} and then analyzed in more 
detail\cite{valenzuela01,grote02,honerkamp02,neumayr03,yamase04b,miyanaga06} 
and also for other models in various subsequent 
works.\cite{metzner03,khavkine04,yamase05a,dellanna06} 
If the $d$FSD tendency is very strong, it can lead to a spontaneous breaking
of the orientational symmetry of the Fermi surface from tetragonal to
orthorhombic. Referring to a stability criterion for Fermi liquids by
Pomeranchuk,\cite{pomeranchuk58} some authors termed this instability 
``Pomeranchuk instability''.

The $d$FSD competes with $d$-wave singlet 
pairing.\cite{yamase00,metzner00,grote02,honerkamp02,neumayr03,yamase04b}   
It differs from other frequently discussed ordering tendencies such as 
antiferromagnetism,\cite{inui88,giamarchi91} 
spin-charge stripes,\cite{tranquada95,white98} staggered flux,\cite{wen96} 
and $d$-density wave,\cite{zeyher99} which are driven by interactions
with a large momentum transfer near $\vq=\vQ$, while the $d$FSD is
generated from forward scattering. 
The $d$FSD breaks the orientational symmetry of the square lattice 
and has the same reduced symmetry as the electronic nematic phase 
proposed by Kivelson, Fradkin and Emery.\cite{kivelson98}
Their route to this phase is however not that of a Fermi surface 
instability, but rather via partial melting of stripe order. 

To discuss high-$T_{c}$ cuprates we should treat both the $d$FSD and 
the $d$-wave singlet pairing on an equal footing. 
We use the slave-boson mean-field scheme of the $t$-$J$ model; 
both tendencies are generated by the $J$ term. 
The $d$FSD tendencies are suppressed by the $d$-wave singlet pairing 
instability such that no spontaneous Fermi surface symmetry breaking 
takes place.\cite{yamase00} 
However, significant $d$FSD correlations survive.\cite{yamase00,yamase04b} 
This correlation effect can drive a sizable enhancement of the Fermi
surface anisotropy by a small bare band anisotropy. 
We analyze this $d$FSD effect on magnetic excitations by using the 
{\it anisotropic} $t$-$J$ model.
The idea that $d$FSD correlations may enhance the in-plane anisotropy 
of magnetic excitations in YBCO has been pursued already in a recent
work by Kao and Kee\cite{kao05}. In that work $d$FSD correlations and
pairing are taken into account via a suitable ansatz for the renormalized 
band structure and the superconducting gap function, while we actually
compute these interaction effects, starting from a microscopic model.

Many features of magnetic excitations in YBCO are captured already by the
(usual) {\it isotropic} $t$-$J$ model.
Magnetic excitations in YBCO are well characterized by the bilayer 
model\cite{tranquada92,mook93,fong9596,dai96,bourges96,bourges98,pailhes03,stock04,pailhes06} 
and therefore have odd and even channels. 
While most neutron scattering studies were confined to the odd 
channel so far, recent experiments successfully detected the even channel 
also.\cite{bourges97,dai99,fong00,pailhes03,pailhes04,stock05,pailhes06}  
We perform a comprehensive analysis of magnetic excitations in the 
bilayer $t$-$J$ model including the above mentioned $d$FSD effects. 
We show that prominent features of magnetic excitations of YBCO 
are well captured in this framework and confirm several theoretical 
insights obtained earlier. The $d$FSD effect turns out to provide 
a natural scenario to understand the observed anisotropy of magnetic 
excitations, especially at low doping and relatively high temperature.  
Together with previous studies of magnetic excitations 
in the single layer $t$-$J$ model\cite{tanamoto94,yamase01,brinckmann01} 
and in the bilayer $t$-$J$ model,\cite{normand95,brinckmann01} 
the present study provides a comprehensive understanding of magnetic 
excitations in YBCO from the fermiology viewpoint. 
Since the antibonding FS can be easily deformed to be open by $d$FSD 
correlations, our work is also relevant for the interpretation of 
ARPES data for untwinned YBCO. 

The article is structured as follows. In Sec.~II, we introduce the bilayer 
$t$-$J$ model and present the slave-boson mean-field scheme. 
In Sec.~III, we discuss self-consistent mean-field solutions of the 
bilayer system. 
Sec.~IV is dedicated to magnetic excitations. After a brief 
qualitative discussion of the dynamical magnetic susceptibility, we 
present numerical results in two parts: (i) the isotropic case, where 
comprehensive results are provided to discuss prominent features of 
magnetic excitations in YBCO, 
and (ii) the anisotropic case, where the $d$FSD effect on magnetic 
excitations is investigated. 
In Sec.~V, we summarize our results through a comparison with 
experimental data, and conclude in Sec. VI.

\section{Model and formalism}

\begin{figure}
\centerline{\includegraphics[width=0.35\textwidth]{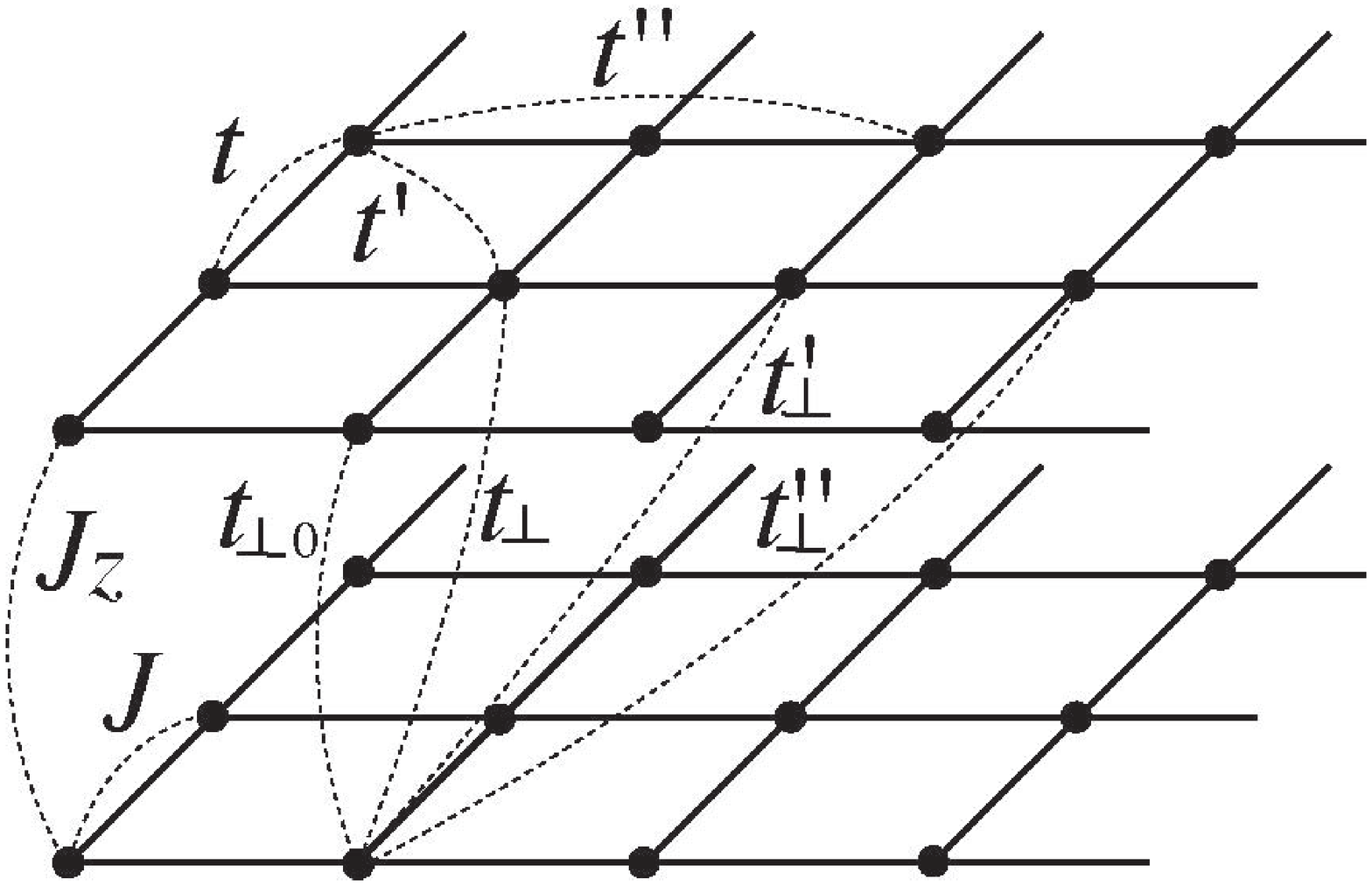}}
\caption{Notation for superexchange couplings 
and transfer integrals in the bilayer square lattice model.}
\label{tJnotation}
\end{figure}

We analyze the bilayer $t$-$J$ model on a square lattice 
\be
 H = -  \sum_{\vr,\,\vr',\, \sigma} t_{\vr\,\vr'} 
 \tilde{c}_{\vr\,\sigma}^{\dagger}\tilde{c}_{\vr'\,\sigma}+
   \sum_{\langle \vr,\vr' \rangle}J_{\vr\,\vr'} \,
 \vS_{\vr} \cdot \vS_{\vr'}  \label{tJ} 
\ee  
defined in the Fock space with no doubly occupied sites.
The operator $\tilde{c}_{\vr\,\sigma}^{\dagger}$
($\tilde{c}_{\vr\,\sigma}$) creates (annihilates) an electron with
spin $\sigma$ on site $\vr$, while $\vS_{\vr}$ is the 
spin operator. 
The site variable $\vr=(x,y,z)$ runs over the bilayer coordinates, 
that is $(x,y)$ indicates a site on the square lattice and 
each layer is denoted by $z=0$ or $1$. 
We assume periodic boundary conditions so that $k_{z}=0$ or $\pi$. 
$J_{\vr\,\vr'}$ $(>0)$ is a superexchange coupling between the 
nearest neighbor sites along each direction, $x,y$, and $z$. 
We take into account hopping amplitudes $t_{\vr\,\vr'}$ between 
$\vr$ and $\vr'$ up to third-nearest neighbors. 
We denote $J_{\vr\,\vr'}$ and $t_{\vr\,\vr'}$ by using the 
conventional notation defined in \fig{tJnotation}. 

We introduce the slave particles, $f_{\vr \sigma}$ and $b_{\vr}$, 
as $\tilde{c}_{\vr \sigma}=b_{\vr}^{\dagger}f_{\vr \sigma}$,  
where $f_{\vr \sigma}$ ($b_{\vr}$) is a fermion (boson) operator 
that carries spin $\sigma$ (charge $e$), and $\vS_{\vr}=\frac{1}{2}
f_{\vr \alpha}^{\dagger}{\bvec \sigma}_{\alpha \beta} f_{\vr \beta}$ 
with the Pauli matrices ${\bvec \sigma} = (\sigma^x,\sigma^y,\sigma^z)$.
The slave bosons and fermions are linked by the local constraint
$b_{\vr}^{\dagger} b_{\vr} + 
 \sum_{\sigma} f_{\vr \sigma}^{\dagger} f_{\vr \sigma} = 1$.
This is an exact transformation known as the slave-boson formalism. 
We then decouple the interaction with the so-called 
resonating-valence-bond (RVB) mean fields:\cite{fukuyama98} 
$\chi_{\boldsymbol \tau}$$\equiv$$\langle \sum_{\sigma}f_{\vr\,\sigma}^{\dagger}
f_{\vr' \,\sigma}\rangle$, 
$\langle b_{\vr}^{\dagger}b_{\vr'}\rangle$, and 
$\Delta_{\boldsymbol \tau}$$\equiv$$\langle f_{\vr\,\uparrow}f_{\vr' \,\downarrow}- 
f_{\vr\,\downarrow}f_{\vr' \,\uparrow}\rangle$, 
with $\bvec{\tau}=\vr'-\vr$. 
These mean fields are assumed to be real constants independent of sites $\vr$. 
We approximate the bosons to condense at the bottom of the band, which 
is reasonable at low temperature and leads to 
$\langle b_{\vr}^{\dagger}b_{\vr'}\rangle=\D$, where $\D$ is the hole 
density. 
The resulting Hamiltonian reads 
\be
\hspace{-0mm}H_{0}=\sum_{\vk}
\left(
      f_{\vk\,\uparrow}^{\dagger}\;\; f_{-\vk\,\downarrow}
\right)
\left( \begin{array}{cc} 
   \xi_{\vk} & -\Delta_{\vk} \\
-\Delta_{\vk} & -\xi_{\vk}
          \end{array}\right)
\left( \begin{array}{c}
 f_{\vk\,\uparrow} \\
 f_{-\vk\,\downarrow}^{\dagger}
\end{array}\right)\, ,\label{MFH}
\ee
with a global constraint 
$\sum_{\sigma}\langle f^{\dagger}_{\vr\sigma}f_{\vr\sigma}\rangle =1-\D$; 
the $\vk$ summation is over $|k_{x(y)}| \leq \pi$, and $k_{z}=0$ and $\pi$. 
Here $\xi_{\vk}=\epsilon_{\vk}^{\|} + \epsilon_{\vk}^{\bot}-\mu$, with the 
in-plane ($c$-axis) dispersion $\epsilon_{\vk}^{\|}$ ($\epsilon_{\vk}^{\bot}$),
and $\Delta_{\vk}=\Delta_{\vk}^{\|}+\Delta_{\vk}{^\bot}$,
with the singlet pairing gap in (out of) the plane $\Delta_{\vk}^{\|}$ 
($\Delta_{\vk}^{\bot}$);
$\mu$ is the chemical potential.
The explicit momentum dependence of the dispersion is given by 
\bea
&&\hspace{-0mm}\epsilon_{\vk}^{\|}=-2\left[ 
  \left(t_{x}\D+\frac{3}{8}J_{x} \chi_{x}\right) \cos k_{x}  
 +\left(t_{y}\D+\frac{3}{8}J_{y} \chi_{y}\right) \cos k_{y}\right.\no\\ 
&& \hspace{-3mm}+ \biggl. 2t'\D \cos k_{x} \cos k_{y}
+ t''_{x}\D \cos 2k_{x}+t''_{y}\D \cos 2k_{y}\biggr]
\label{inxi}\; ,\\
&&\hspace{-0mm}\epsilon_{\vk}^{\bot}=-\left[
\left(t_{\bot 0}\D+\frac{3}{8}J_{z}\chi_{z}\right)
+2 \left(t_{\bot x}\D \cos k_{x} + t_{\bot y}\D \cos k_{y}\right.
\right.\no\\ 
&& \hspace{-3mm}+ \biggl.\left.  2t'_{\bot}\D \cos k_{x} \cos k_{y} 
+ t''_{\bot x}\D \cos 2k_{x}+t''_{\bot y}\D \cos 2k_{y}\right)
\biggr]\cos k_{z}  
\label{outxi}\; ,
\eea
and that of the gap function by
\bea
&&\hspace{-0mm}\Delta_{\vk}^{\|}=-\frac{3}{4}  
 \left(J_{x}\Delta_{x}\cos k_{x}+J_{y}\Delta_{y} \cos k_{y}\right)
\label{insc} \,, \\
&&\hspace{-0mm}\Delta_{\vk}^{\bot}=-\frac{3}{8}J_{z}\Delta_{z}\cos k_{z}
\label{outsc} \,. 
\eea

In the bilayer high-$T_{c}$ cuprates, 
a $(\cos k_{x}-\cos k_{y})^{2}$-type $c$-axis dispersion was 
computed\cite{andersen95} and actually observed in 
Bi$_{2}$Sr$_{2}$CaCu$_{2}$O$_{8+\D}$.\cite{feng01,chuang01} 
To adapt \eq{outxi} to this behavior, we choose transverse hopping
amplitudes as follows: 
$t_{\bot 0}=\frac{\gamma_{x}^{2}+\gamma_{y}^{2}}{2}t_{z}$, 
$t_{\bot x}=t_{\bot y}=0$, 
$t_{\bot}'=-\frac{t_{z}\gamma_{x}\gamma_{y}}{2}$, 
$t_{\bot x}''=\frac{\gamma_{x}^{2}}{4}t_{z}$, and 
$t_{\bot y}''=\frac{\gamma_{y}^{2}}{4}t_{z}$.
The parameters $\gamma_{x}$ and $\gamma_{y}$ allow for a convenient 
parametrization of an in-plane anisotropy as specified below.
The resulting $c$-axis dispersion reads 
\be
\epsilon_{\vk}^{\bot}=-\left[t_{z}\D\left(\gamma_{x}\cos k_{x}
-\gamma_{y}\cos k_{y}\right)^{2}+\frac{3}{8}J_{z}\chi_{z}\right]
\cos k_{z}\,,\label{caxis}
\ee
which has the expected form in the isotropic case 
$\gamma_{x}=\gamma_{y}=1$. 

YBa$_{2}$Cu$_{3}$O$_{y}$ has an orthorhombic crystal structure for 
$y\geq 6.4$, where the superconducting state is realized at low 
$T$.\cite{jorgensen90} 
Such orthorhombicity yields $xy$-anisotropy, which we incorporate by 
introducing a single parameter $\alpha$ as follows: 
\bea
&&t_{x}=t(1+\alpha/2),\quad t_{y}=t(1-\alpha/2),\\
&&J_{x}=J(1+\alpha),\quad J_{y}=J(1-\alpha),\\
&&t_{x}''=t''(1+\alpha/2),\quad t_{y}''=t''(1-\alpha/2),\\
&&\gamma_{x}=1+\alpha/4,\quad \gamma_{y}=1-\alpha/4\,.
\eea
With this parametrization, $t_{x(y)}$, $t''_{x(y)}$, and $t''_{\perp x(y)}$ 
have the same degree of $xy$-anisotropy, 
while the anisotropy of $J_{x(y)}$ is twice as large, as imposed by
the superexchange mechanism. 

We determine $\chi_{\tau}$ and $\Delta_{\tau}$ with $\tau=x,y,z$ 
by solving the following self-consistency equations numerically: 
\bea
&&\chi_{\tau}=-\frac{1}{N}\sum_{\vk}\cos k_{\tau}\frac{\xi_{\vk}}{E_{\vk}} 
\tanh \frac{E_{\vk}}{2T}\,,\\
&&\Delta_{\tau}=-\frac{1}{N}\sum_{\vk}\cos k_{\tau}
\frac{\Delta_{\vk}}{E_{\vk}} \tanh \frac{E_{\vk}}{2T}\,,\\
&&\D=\frac{1}{N}\sum_{\vk}\frac{\xi_{\vk}}{E_{\vk}} 
\tanh \frac{E_{\vk}}{2T}\,,
\eea
where $E_{\vk}=\sqrt{\xi_{\vk}^{2}+\Delta_{\vk}^{2}}$ and 
$N$ is the total number of (bilayer) lattice sites.

\section{Self-consistent solution}
 
The material dependence of high-$T_{c}$ cuprates 
is mainly taken into account by different choices of band 
parameters.\cite{tanamoto93,feiner96,tohyama00,pavarini01} 
We use the following parameters for YBCO: 
\be
t/J=2.5,\,t'/t=-0.3,\,t''/t=0.15,\,J_{z}/J=0.1,\, t_{z}/t=0.15\,, 
{\rm and}\ \alpha = -0.05\,. \no
\ee
This choice has been done judiciously.
{\it Ab initio} calculations\cite{hybertsen90} indicate that realistic 
values for $t/J$ lie in the range $2-5$.  
Even within this restricted interval, we found that the behavior 
of Im$\chi(\vq,\,\omega)$ strongly depends on the choice of $t/J$. 
We have chosen $t/J$ in such a way that the energy of the resonance 
mode at $\vq = (\pi,\pi)$ is maximal at optimal doping, that
is near $\delta=0.15$ (see \fig{d-wQ}), to roughly agree with 
experiments.\cite{fong00,dai01,stock04,pailhes06}  
The ratios $t'/t$ and $t''/t$ are extracted from an LDA band 
calculation for YBCO.\cite{andersen95}  
The value of $J_{z}/J$ is fixed rather uniquely by the optical 
magnon energy in YBCO.\cite{hayden96,reznik96} 
The bilayer system has two FSs, the bonding band FS and 
the antibonding band FS. 
With increasing $\D$, the latter can change from a hole-like FS 
to an electron-like FS in the region we are interested in. 
The hole density at which such a FS topology change occurs strongly 
depends on $t_{z}$. 
With our choice of $t_z/t$, which is a bit larger than LDA 
estimates,\cite{andersen95} the antibonding FS becomes electron-like 
for $\D\gtrsim 0.20$.
Some ARPES data for Bi$_{2}$Sr$_{2}$CaCu$_{2}$O$_{8+\D}$ indeed 
suggest an electron-like antibonding FS for sufficiently high 
overdoping.\cite{chuang01,bogdanov01} 
ARPES data for YBCO with such a high doping are not yet available. 

It is not easy to determine the anisotropy parameter $\alpha$; 
even its sign is not obvious. 
The orthorhombic YBCO has a lattice constant anisotropy $a < b$, 
where $a (b)$ is the lattice constant along the $x (y)$ direction. 
This contributes to an enhancement of $t_{x}$. 
On the other hand, YBCO contains CuO chains along the $y$ direction. 
A small hybridization with the chain band enhances $t_{y}$, an 
effect opposite to the lattice constant effect. 
If one estimates a band anisotropy from the predicted FSs in 
Ref.~\cite{andersen95}, a small negative $\alpha$ is obtained. 
A most recent LDA calculation supports this estimate and predicts 
that $t_{y}$ is larger than $t_{x}$ by about 
$3-4\%$.\cite{andersen05} 
Hence we assume $t_{x} < t_{y}$ and set $\alpha=-0.05$ when 
analyzing the in-plane anisotropy of Im$\chi(\vq,\,\omega)$.
We will also present results for the isotropic case, $\alpha=0$, 
for comparison.

We introduce the following convenient notation:
\bea
&&\chi_{0}=\frac{\chi_{x}+\chi_{y}}{2},\quad
\chi_{d}=\frac{\chi_{x}-\chi_{y}}{2},  \\
&&\Delta_{d}=\frac{\Delta_{x}-\Delta_{y}}{2},\quad 
\Delta_{s}=\frac{\Delta_{x}+\Delta_{y}}{2}\,, 
\eea
where $\chi_{d}$ is an order parameter of the $d$FSD, 
and $\Delta_{d} (\Delta_{s})$ is the $d$-wave ($s$-wave) 
in-plane singlet pairing amplitude; we choose $\Delta_{x}\geq 0$. 

\begin{figure}
\centerline{\includegraphics[width=0.35\textwidth]{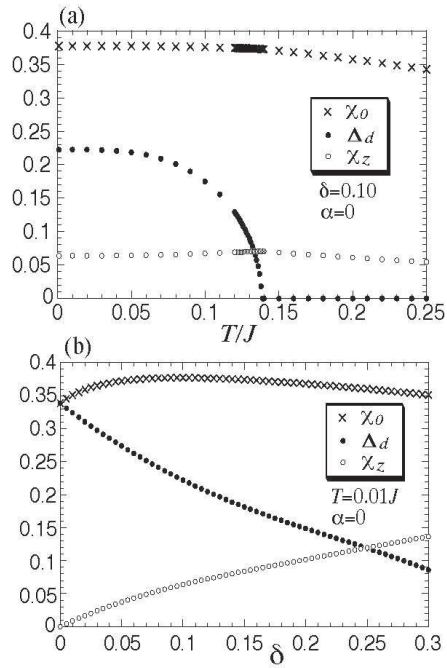}}
\caption{Mean fields $\chi_{0}$, $\Delta_{d}$, and $\chi_{z}$  
for $\alpha=0$: (a) $T$ dependence for $\D=0.10$, and 
(b) $\D$ dependence at $T=0.01J$. 
}
\label{meana0}
\end{figure}

For $\alpha=0$, there is no $d$FSD,\cite{yamase00,yamase04b}
that is, $\chi_d = 0$, and an isotropic $d$-wave pairing state is 
stabilized at low $T$, that is, $\Delta_{s}=\Delta_{z}=0$. 
Note that a finite $\Delta_{z}$ contributes to an $s$-wave pairing 
component of $\Delta_{\vk}$ [see \eq{outsc}]. 
The temperature dependences of $\chi_{0}$, $\chi_{z}$, and $\Delta_{d}$ 
are shown in \fig{meana0}(a) and their $\delta$ dependences in 
\fig{meana0}(b). 
$\chi_{0}$ and $\chi_{z}$ depend very weakly on $T$. 
In particular, the onset of $\Delta_{d}$ does not affect 
$\chi_{0}$ and $\chi_{z}$ substantially. 
At $\delta=0$, $\chi_{z}$ becomes zero, that is, the layers are 
decoupled regardless of the interlayer coupling $J_{z}$ in the present
mean-field framework, where antiferromagnetic 
order is not taken into account.  
The layers couple weakly only for finite $\delta$. 
The $d$-wave pairing amplitude $\Delta_{d}$ increases with decreasing 
$\delta$, which however should not be interpreted as an enhancement of 
a superconducting order parameter at lower $\delta$, but rather as the 
increase of a one-particle pseudogap energy scale in the slave-boson 
mean-field scheme.\cite{fukuyama98} 

\begin{figure}
\centerline{\includegraphics[width=0.45\textwidth]{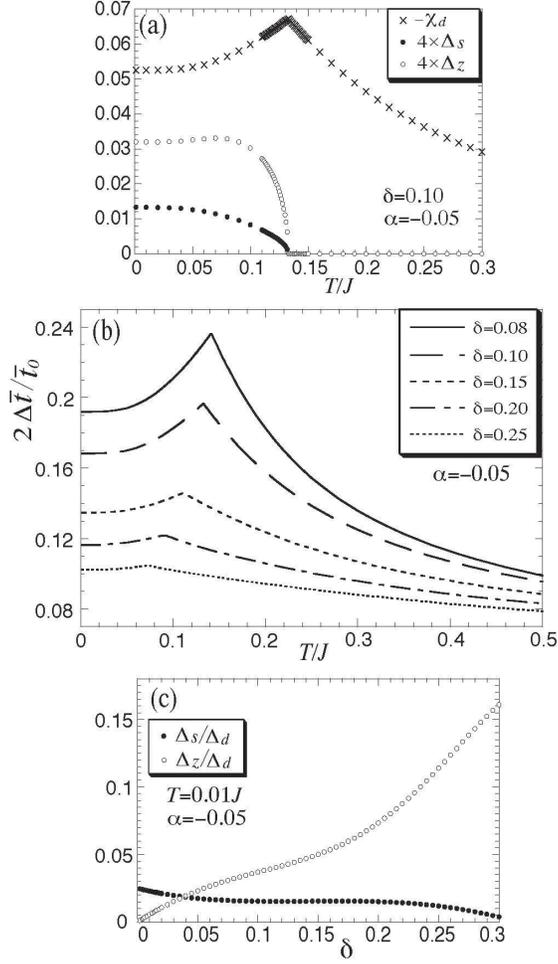}}
\caption{(a) $T$ dependence of $\chi_{d}$, $\Delta_{s}$, and $\Delta_{z}$ 
for $\D=0.10$ and $\alpha=-0.05$; 
$\Delta_{s}$ and $\Delta_{z}$ are multiplied by 4; note that 
$\chi_{d}<0$ because of $\alpha<0$.  
(b) $T$ dependence of anisotropy of the renormalized band, 
$2\Delta \bar t /\Delta \bar t_{0}$, for several choices of $\D$. 
(c) $\D$ dependence of $\Delta_{s}$ and $\Delta_{z}$ at $T=0.01J$. 
}
\label{meana5}
\end{figure}

When the anisotropy parameter $\alpha$ is introduced, the system loses 
tetragonal symmetry, and $\chi_{d}$, $\Delta_{s}$, and $\Delta_{z}$ 
can become finite. 
Since the temperature and doping dependences of $\chi_{0}$, 
$\Delta_{d}$, and $\chi_{z}$ for $\alpha=-0.05$ are almost the same 
as in \fig{meana0}, we focus on $\chi_{d}$, $\Delta_{s}$, and 
$\Delta_{z}$. 
The temperature dependences of $\chi_{d}$, $\Delta_{s}$ and $\Delta_{z}$ 
are shown in \fig{meana5}(a). 
$\chi_{d}$ increases with decreasing $T$, which is due to the 
development of $d$FSD correlations, and exhibits a cusp at the onset of 
singlet pairing, which is denoted as $T_{\rm RVB}$ in the slave-boson 
theory; note that $\Delta_{d}$, $\Delta_{z}$, and $\Delta_{s}$ have the 
same onset temperature.
Below $T_{\rm RVB}$, $d$FSD correlations are suppressed leading to a 
suppression of $\chi_{d}$. 
But $\chi_{d}$ saturates at a finite value at zero temperature and is 
still enhanced compared to $\chi_{d}$ at high $T$. 
Since the renormalized band is mainly characterized by the renormalized
nearest-neighbor hopping 
$\bar t_{\tau} = t_{\tau}\D + \frac{3}{8}J_{\tau}\chi_{\tau}$ 
[\eq{inxi}], this enhancement contributes to the band anisotropy. 
Defining $\Delta \bar t = |\bar t_x - \bar t_y|/2$ and 
$\bar t_0 = (\bar t_x + \bar t_y)/2$, we plot 
$2\Delta \bar t/\bar t_0$ in \fig{meana5}(b) for several choices of 
$\delta$. Although the bare anisotropy is only $5\%$, the renormalized
band anisotropy is enhanced up to around $25\%$ for $\delta = 0.08$ 
by $d$FSD correlations. 
The enhancement is less prominent at higher $\delta$. 

Figure~\ref{meana5}(c) shows the doping dependences of $\Delta_{s}$ and 
$\Delta_{z}$ at low $T$.  
Although $\Delta_{s}$ increases at lower $\delta$, its magnitude 
remains very small compared to $\Delta_{d}$. 
Note that we obtain a positive $\Delta_{s}$ for $\alpha=-0.05$, 
that is $\Delta_{x} > |\Delta_{y}|$ for $J_{x} < J_{y}$, which 
means that the $s$-wave component reduces the anisotropy of 
$|\Delta_{\vk}^{\|}| \,$ caused by the anisotropy of $J$ [see \eq{insc}]. 
The out-of-plane pairing $\Delta_{z}$ becomes finite at finite $\delta$. 
Its sign is determined uniquely through the linear coupling to 
$\Delta_{d}$. 
It becomes positive for the present band parameters. 
In the antibonding band ($k_{z}=\pi$), $\Delta_{z}$ enhances the 
magnitude of $\Delta_{\vk}$ at $(\pi,\,0)$ and reduces it at $(0,\,\pi)$,
and vice versa in the bonding band ($k_{z}=0$). 

\begin{figure}
\centerline{\includegraphics[width=0.45\textwidth]{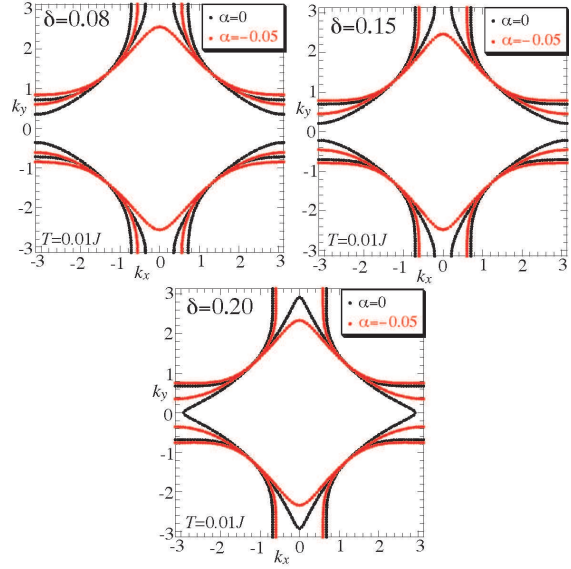}}
\caption{(color online) 
Fermi surfaces for several choices of $\D$ at low $T$ for 
$\alpha=0$ (black lines) and $\alpha=-0.05$ (red lines).}
\label{FS}
\end{figure}

\begin{figure}
\centerline{\includegraphics[width=0.40\textwidth]{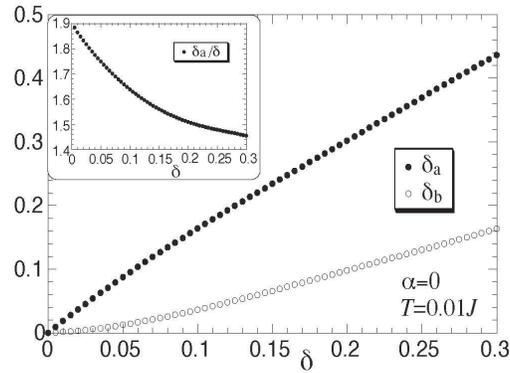}}
\caption{Hole density in the antibonding band ($\D_{a}$) 
and the bonding band ($\D_{b}$) as a function of 
total hole density $\D$; note that $\D=(\D_{a}+\D_{b})/2$. 
The inset shows the ratio $\D_{a}/\D$.}
\label{density}
\end{figure}

The FS at low $T$, which we define by $\xi_{\vk}=0$ in the singlet 
pairing state, is shown in Figs.~\ref{FS}(a)-(c) 
for several choices of $\delta$. 
The bonding band ($k_{z}=0$) forms the outer FS 
and the antibonding band ($k_{z}=\pi$) the inner FS. 
For $\alpha=0$, the inner FS changes into an electron-like FS 
for $\D\gtrsim 0.20$ while the outer FS stays hole-like. 
For $\alpha=-0.05$, the inner FS becomes open for all $\delta$ in 
\fig{FS}. 
The inner FS always opens already for a tiny anisotropy parameter 
for doping around $\delta \approx 0.20$, since the original inner 
FS is close to $(\pi,\,0)$ and $(0,\,\pi)$ in this case. 
It depends sensitively on the size of $\alpha$ whether an open FS 
is realized at low $\D$ where $d$FSD correlations become large but 
the original inner FS is away from $(\pi,\,0)$ and $(0,\,\pi)$.
The antibonding band and the bonding band have different hole 
densities, $\delta_{a}$ and $\delta_{b}$, respectively, 
as shown in \fig{density}. From the inset of \fig{density} 
we see that the antibonding band has a hole density more than 1.5 
times higher than the total (average) hole density $\delta$.

\section{Magnetic excitations}

Next we investigate magnetic excitation spectra in the bilayer 
$t$-$J$ model using the self-consistent mean-field solutions. 
The irreducible dynamical magnetic susceptibility 
$\chi_{0}(\vq,\,\omega)$ is given by 
\bea
\chi_{0}(\vq,\,\omega) &=& \frac{1}{4 N} 
\sum_{\vk}\left[C^{+}_{\vk,\,\vk+\vq} 
 \left(\tanh \frac{E_{\vk}}{2T}
   -\tanh \frac{E_{\vk +\vq}}{2T}\right)
   \frac{1}{E_{\vk}-E_{\vk+\vq}+\omega+{\rm i}\Gamma}\right. \no\\
 &+& \frac{1}{2} \, C^{-}_{\vk,\,\vk+\vq}
\left(\tanh \frac{E_{\vk}}{2T}
   +\tanh \frac{E_{\vk +\vq}}{2T}\right) \no\\
 && \times \left. \left(\frac{1}{E_{\vk}+E_{\vk+\vq}+\omega+{\rm i}\Gamma}
   +\frac{1}{E_{\vk}+E_{\vk+\vq}-\omega-{\rm i}\Gamma}\right)\right],\,
\label{xoqw}
\eea
where $\Gamma$ is a positive infinitesimal and 
\be
C^{\pm}_{\vk,\,\vk+\vq}=\frac{1}{2}
\left(1 \pm \frac{\xi_{\vk}\xi_{\vk+\vq}
 +\Delta_{\vk}\Delta_{\vk+\vq}}{E_{\vk}E_{\vk+\vq}}\right)\; .
\ee
Note that the $k_z$-component of $\vk$ is summed only over the two values
$k_{z}=0$ and $\pi$, corresponding to a bonding and an antibonding band,
respectively. 
Particle-hole scattering processes are therefore intraband processes for 
$q_{z}=0$ and interband processes for $q_{z}=\pi$. 
The former is called the even channel and the latter the odd channel. 

In a renormalized random phase approximation (RPA)\cite{brinckmann99,yamase99} 
the dynamical magnetic susceptibility $\chi(\vq,\,\omega)$ is given by 
\be
\chi(\vq,\,\omega)=\frac{\chi_{0}(\vq,\,\omega)}{1+ 
J(\vq)\chi_{0}(\vq,\,\omega)}\; , \label{RPA}
\ee
where 
\be
J(\vq) = 2r (J_{x}\cos q_{x}+J_{y}\cos q_{y}) +J_{z}\cos q_{z} \label{RPAJ}
\ee
with a renormalization factor $r$.
In the 
plain RPA one has $r=1$, which leads to a divergence of $\chi(\vq,\,0)$ 
at $\vq \approx (\pi,\,\pi,\,\pi)$ in a wide doping region 
$(\delta \lesssim 0.20)$, signaling an instability toward the AF 
state. 
However, several numerical studies of the $t$-$J$ model indicate that 
the antiferromagnetic instability is overestimated by the 
RPA.\cite{chen90,giamarchi91,himeda99} 
Fluctuations not included in RPA obviously suppress the instability.
This can be taken into account in a rough and phenomenological way by 
setting $r<1$.\cite{brinckmann99,yamase99}
Here we choose the value $r = 0.5$, which confines the AF instability 
to $\delta \leq 0.064$, consistent with actual YBCO 
samples.\cite{jorgensen90,akoshima98} 

In the following we will frequently specify $q_z$ by refering to the
``even'' or ``odd'' channel, and include only the remaining $q_x$ and
$q_y$ variables in $\vq$.

\subsection{Basic properties} 

\begin{figure}
\centerline{\includegraphics[width=0.45\textwidth]{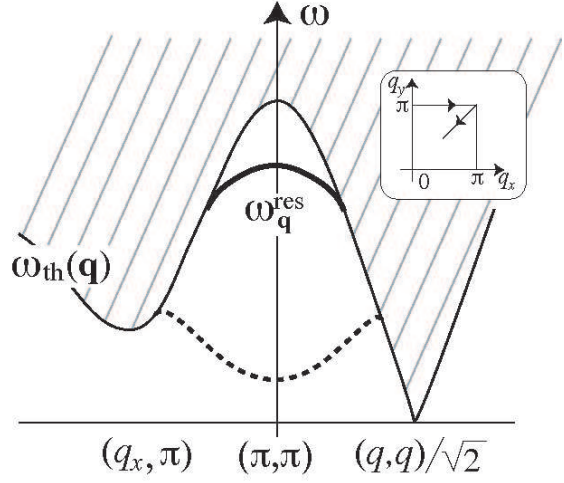}}
\caption{Schematic plot of magnetic excitation spectra 
Im$\chi(\vq,\,\omega)$; the $\vq$ directions are shown in the inset. 
The hatched region represents the continuum spectrum and its lower edge 
is $\omega_{\rm th}(\vq)$. The bold solid line $\omega_{\vq}^{\rm res}$ 
represents a typical dispersion of the collective mode. 
The mode is softened near the AF instability  leading to an upward
dispersion as shown by the dotted line.}
\label{wthfig}
\end{figure}

Before calculating the dynamical magnetic susceptibility numerically, 
we first discuss generic properties of Im$\chi(\vq,\,\omega)$, which 
hold in both odd and even channels. 
At zero temperature the $d$-wave superconducting state is realized and 
\be
{\rm Im}\chi_{0}(\vq,\,\omega)=\frac{\pi}{4N}
\sum_{\vk}C^{-}_{\vk,\vk+\vq} \D(E_{\vk}+E_{\vk+\vq}-\omega)\,.
\ee
Hence, Im$\chi_{0}(\vq,\,\omega)$ has a threshold energy defined by 
\be
\omega_{\rm th}(\vq)={\rm min}\{E_{\vk}+E_{\vk+\vq},\,\vk \in {\rm BZ}\}\, ,
\label{wth}
\ee
above which continuum excitations start. 
$\omega_{\rm th}(\vq)$ is sketched by the solid line in \fig{wthfig} 
and has a maximum at $\vq=\vQ$.  
Along the diagonal direction $\vq=\frac{1}{\sqrt{2}}(q,\,q)$, gapless 
excitations appear, which are due to scattering processes between the 
$d$-wave gap nodes. 

The denominator in \eq{RPA} vanishes if both real and imaginary parts
vanish, that is 
\bea
&&1+J(\vq){\rm Re}\chi_{0}(\vq,\,\omega_{\vq}^{\rm res})=0,\label{wq1}\\
&&{\rm Im}\chi_{0}(\vq,\,\omega_{\vq}^{\rm res})=0 \label{wq2}\,.
\eea 
As widely discussed in the literature,\cite{bulut96,brinckmann99,norman00,
brinckmann01,chubukov01,li02,onufrieva02,schnyder04,zhou0405} 
Eqs.~(\ref{wq1}) and (\ref{wq2}) can be simultaneously satisfied 
when Im$\chi_{0}(\vq,\,\omega)$ jumps from zero to a finite
value at $\omega=\omega_{\rm th}(\vq)$, which gives rise to a 
logarithmic divergence of Re$\chi_{0}(\vq,\,\omega)$,
such that there is always a solution 
$\omega=\omega_{\vq}^{\rm res}<\omega_{\rm th}(\vq)$. 
This solution describes an ingap collective mode of particle-hole 
excitations with spin 1 and charge zero. 
Note that $\vq$ is not restricted to $\vQ$. In fact, the collective 
mode appears also around $\vq=\vQ \,$, in particular at IC or DIC 
wavevectors as sketched in \fig{wthfig}.   
Expanding $J(\vq){\rm Re}\chi_{0}(\vq,\,\omega)$ around $\vq=\vQ$ and 
$\omega=\omega_{\vQ}^{\rm res}$, 
we obtain the asymptotic form of the dispersion 
\be
\omega_{\vq}^{\rm res}=\omega_{\vQ}^{\rm res}
-\frac{1}{2}\left[\frac{(q_{x}-\pi)^{2}}{m_{x}}+
\frac{(q_{y}-\pi)^{2}}{m_{y}}\right]
\,, \label{wq}
\ee
where
\be
m^{-1}_{x(y)} = \left. \left. [J(\vQ)]^{-1} \, 
\frac{\partial^{2}J(\vq)
{\rm Re}\chi_{0}(\vq,\,\omega)}{\partial {q}_{x(y)}^{2}}
\right/\frac{\partial {\rm Re}\chi_{0}(\vq,\,\omega)}{\partial\omega}
\, \right|_{\, \vq=\vQ , \, \omega = \omega_{\vQ}^{\rm res}} \; . 
\label{wqm}
\ee
The denominator in \eq{wqm} 
is usually positive for $\omega <\omega_{\rm th}(\vQ)$.  
The numerator is expected to be negative close to the commensurate 
AF instability since $|J(\vq){\rm Re}\chi_{0}(\vq,\,\omega)|$ 
has a commensurate peak and $J(\vq)<0$, leading to an upward dispersion 
(dotted line in \fig{wthfig}). 
With hole doping, the peak of  $|J(\vq){\rm Re}\chi_{0}(\vq,\,\omega)|$  
usually shifts to IC positions, and thus its curvature at $\vq=\vQ$ 
becomes positive, that is, $\omega_{\vq}^{\rm res}$ has a downward 
dispersion (bold solid line in \fig{wthfig}). 
When the collective mode mixes with the continuum, it disappears
due to overdamping. 

The infinitesimal $\Gamma$ in \eq{xoqw} is 
replaced by a finite small value in numerical calculations, 
leading to a cutoff of the logarithmic divergence of 
Re$\chi_{0}(\vq,\,\omega)$. A similar cutoff occurs also for 
finite $T$. 
Then \eq{wq1} can be satisfied only when Im$\chi_{0}(\vq,\,\omega)$ 
has a sufficiently large jump at $\omega=\omega_{\rm th}(\vq)$ or when
Re$\chi_{0}(\vq,\,\omega)$ is sufficiently large; the latter is always 
the case near the AF instability. 
Practically therefore the collective mode is well defined only in a 
limited momentum region (\fig{wthfig}).   
Moreover, since Im$\chi_{0}(\vq,\,\omega)$ becomes finite for a finite 
$\Gamma$ or a finite $T$, the collective mode has a finite life time, 
which yields a finite width for the peak in Im$\chi(\vq,\,\omega)$.

\subsection{Numerical calculation}

Now we perform extensive numerical calculations of Im$\chi(\vq,\,\omega)$
with a small finite $\Gamma=0.01J$ in the denominator of \eq{xoqw}. 
Although the choice of a finite $\Gamma$ is mainly done for numerical 
convenience, it also simulates damping of electrons by static defects in 
real materials, and broadening due to limited energy resolution in 
inelastic neutron scattering experiments. 

This section is composed of six parts, where the first three deal with 
the isotropic case ($\alpha=0$) and the last three with the effects due 
to anisotropy ($\alpha\neq 0$). 
The former are intended to give a comprehensive analysis of magnetic
excitations in bilayer cuprates such as YBCO, without taking effects 
which are specifically due to in-plane anisotropy into account: 
1. $(\vq,\,\omega)$ maps of Im$\chi(\vq,\,\omega)$  for a sequence of 
$\delta$ for both odd and even channels, 
2. $\vq$ maps for a sequence of $\omega$, and 
3. $T$ dependence. 
The latter are dedicated to anisotropy effects on magnetic excitations
and their enhancement by $d$FSD correlations: 
4. $\vq$ maps for a sequence of $\omega$, 
5. $\vq$ maps for several choices of $\delta$, and 
6. $\vq$ maps for a sequence of temperatures.

\subsubsection{$(\vq,\,\omega)$ maps for $\alpha=0$}
 
\begin{figure}
\centerline{\includegraphics[width=0.5\textwidth]{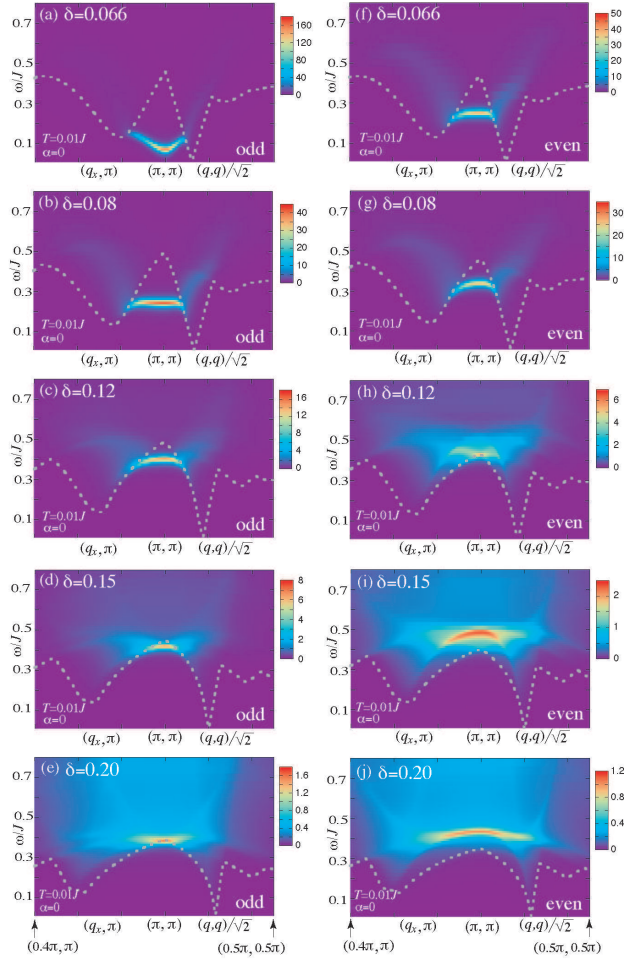}}
\caption{(color) 
$(\vq,\,\omega)$ maps of Im$\chi(\vq,\,\omega)$ for a sequence of 
doping concentrations $\D$ at $T=0.01J$ and $\alpha=0$ 
for both odd (left panels) and even (right panels) channels; 
the gray dotted line is the threshold energy $\omega_{\rm th}(\vq)$; 
the $\vq$ scan directions are shown in the inset of \fig{wthfig}: 
$(0.4\pi,\,\pi) \leq \vq \leq (\pi,\,\pi)$ and 
$(\pi,\,\pi) \geq \vq \geq (0.5\pi,\,0.5\pi)$. 
}
\label{qwa0-d}
\end{figure}

Figure~\ref{qwa0-d} shows intensity maps of {\rm Im}$\chi(\vq,\,\omega)$ 
for a sequence of hole densities $\delta$ for the odd channel (left panels) 
and the even channel (right panels); the $\vq$ scan direction is shown 
in the inset of \fig{wthfig}. 
The threshold energy $\omega_{\rm th}(\vq)$ [\eq{wth}] is also plotted 
with a gray dotted line. 
Figure~\ref{qwa0-d}(a) shows a result in the vicinity of the AF instability. 
Strong intensity is seen well below $\omega_{\rm th}(\vq)$, indicating 
collective particle-hole excitations with the upward dispersion described 
by \eq{wq}. 
This soft collective mode reflects the commensurate AF instability 
at $\delta \approx 0.064$. 
The mode disperses as a function of $\vq$ and mixes into the particle-hole 
continuum at sufficiently large distance from $\vQ$, where it becomes
overdamped and thus faintly visible on the color scale in \fig{qwa0-d}(a). 
When $\delta$ is slightly increased, the soft collective mode rapidly
shifts to higher energies. It has a nearly flat dispersion around 
$\vq=\vQ$ for $\delta=0.08$ [\fig{qwa0-d}(b)] and a clear downward 
dispersion for $\delta=0.12$ [\fig{qwa0-d}(c)]. 
In the particle-hole continuum above $\omega_{\rm th}(\vq)$, significant 
intensity is seen in Figs.~\ref{qwa0-d}(b) and (c) with an upward dispersion. 
This can be interpreted as an overdamped collective mode near 
$\omega_{\vQ}^{\rm res}$, as we shall discuss later. 
With further increasing $\delta$, the collective mode is pushed up to 
higher energy and appears only close to $\omega=\omega_{\rm th}(\vQ)$ 
[\fig{qwa0-d}(d)]. 
Above $\omega_{\rm th}(\vQ)$, the spectrum broadens and the structures 
become less clear. 
For $\delta=0.20$ [\fig{qwa0-d}(e)], the collective mode does not 
appear and we just see that the continuum spectrum has strong intensity 
near $\vq=\vQ$ and $\omega=\omega_{\rm th}(\vQ)$, and a very broad 
featureless distribution above. 
As a function of $\D$, Figs.~\ref{qwa0-d}(a)-(e) show that the spectral 
weight becomes larger at lower $\D$ (see values of the color map index) 
due to the proximity to the AF instability. 

The corresponding results for the even channel are shown 
in the right panels in \fig{qwa0-d}. 
The overall features look the same as those of the odd channel. 
Close to the AF instability [\fig{qwa0-d}(f)], however, the 
collective mode retains the downward dispersion and does not show 
softening, reflecting the fact that the AF instability is driven 
by the odd channel. 
The collective mode is pushed up to higher energy with $\D$ 
[Figs.~\ref{qwa0-d}(g) and (h)] and appears inside the continuum 
in \fig{qwa0-d}(i) in the sense that \eq{wq1} is still satisfied. 
The mode in the continuum is substantially overdamped but has 
the same $|\vq-\vQ|^{2}$ dispersion as \eq{wq}. 
The coefficient, however, is not given by \eq{wqm}, which is valid
only if ${\rm Im}\chi_0(\vq,\omega_{\vq}^{\rm res}) = 0$. 
Figure~\ref{qwa0-d}(j) looks similar to \fig{qwa0-d}(i), but \eq{wq1} 
is not satisfied any longer and the strong intensity results from 
individual excitations. 
Compared to the odd channel, we see that spectral weight of the even 
channel is overall suppressed and broadened.  
This difference comes from two effects:  
(i) the suppression of $|J(\vQ)|$ for the  
even channel near $\vq=\vQ$ [see Eq.~(\ref{RPAJ})], 
and (ii) different particle-hole scattering processes, 
namely interband scatterings for the odd channel and 
the intraband scatterings for the even channel. 
The former (latter) effect is dominant for low (high) $\D$. 

To see the intensity profile of Im$\chi(\vq,\,\omega)$ more clearly, 
we replot \fig{qwa0-d}(c) by separating it into three energy regions 
in the left panel in Fig.~\ref{qwdetail} with an optimized color map 
scale in each;  
the right panel is a map of $|1+J(\vq){\rm Re}\chi_{0}(\vq,\,\omega)|$, 
which serves to quantify the collective character of magnetic 
excitations [see \eq{wq1}]; the cross symbols indicate peak intensity 
positions of Im$\chi(\vq,\,\omega)$.
For $\omega\lesssim \omega_{\vQ}^{\rm res}=0.403J$, the ingap collective 
mode has a downward dispersion. 
This collective mode is however realized only in a limited energy 
region. 
See the region where $|1+J(\vq){\rm Re}\chi_{0}(\vq,\,\omega)|$ 
becomes zero in the right panel;
at lower $\omega$, $|1+J(\vq){\rm Re}\chi_{0}(\vq,\,\omega)|$ remains
finite such that the peak in Im$\chi(\vq,\omega)$ does not correspond
to a genuine collective mode.
With further decreasing $\omega$ the IC signals are substantially 
diminished, while the DIC signals continue down to zero energy. 
For $\omega>\omega_{\vQ}^{\rm res}$, we see another dispersive 
structure inside the continuum. 
This upward dispersion can be interpreted as an overdamped collective 
mode up to $\omega\sim 0.50J$ in the sense that
$|1+J(\vq){\rm Re}\chi_{0}(\vq,\,\omega)|$ becomes zero near 
the peak position. 
Note that this overdamped mode is not directly connected with the 
downward collective mode below $\omega_{\vQ}^{\rm res}$. 
Although the overdamped mode seems to continue smoothly 
up to higher energy in the left panel, it changes into a peak of 
individual excitations as seen from the finite value of 
$|1+J(\vq){\rm Re}\chi_{0}(\vq,\,\omega)|$ in the right panel. 

\begin{figure}
\centerline{\includegraphics[width=0.35\textwidth]{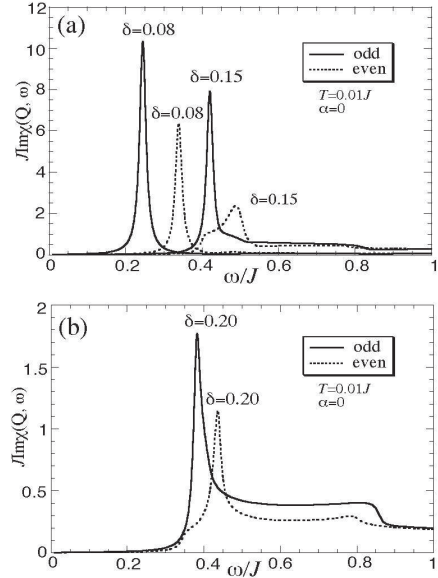}}
\caption{$\omega$ dependence of Im$\chi(\vQ,\,\omega)$ for 
$\D=0.08$ and $0.15$ (a), and $\D=0.20$ (b)
for the odd and even channels; the results for $\D=0.08$ are multiplied 
by 0.2. 
}
\label{wa0-d}
\end{figure}

In \fig{wa0-d}, we plot Im$\chi(\vq,\omega)$ at $\vq=\vQ$ as a 
function of $\omega$ for (a) $\D=0.08$ and $0.15$, and (b) $\D=0.20$. 
The sharp peak for $\D=0.08$ corresponds to the collective mode;
the continuum cannot be seen on the scale of \fig{wa0-d}(a).  
For $\D=0.15$ in the odd channel, we see both the collective mode 
at $\omega =0.42J$ and the continuum above $\omega=0.45J$; the two 
are not completely separated because of finite $\Gamma$ and $T$. 
The peak at $\omega=0.49J$ for the even channel is a collective 
mode, but its peak width is much broader than that for the odd 
channel, indicating a much shorter life time of the mode.  
The odd channel spectrum for $\D=0.20$ is similar to that for $\D=0.15$. 
The peak at $\omega=0.38J$, however, is not a collective mode;  
Eq.~(\ref{wq1}) is not satisfied any longer. 
Similarly the peak for the even channel results just from 
individual excitations. 

We summarize the peak position of Im$\chi(\vq,\,\omega)$ at 
$\vq=\vQ$, which we refer to as $\omega=\omega_{\vQ}$, 
as a function of $\D$ in \fig{d-wQ}; $\omega_{\vQ}$ is equivalent
to $\omega_{\vQ}^{\rm res}$ when \eq{wq1} is satisfied; 
the lower edge energy of particle-hole continuum, 
$\omega_{\rm th}(\vQ)$, is also plotted by the bold gray lines; 
we consider the region, $\D \geq 0.064$, below which the AF 
instability takes place.  
With increasing doping, $\omega_{\vQ}$ increases rapidly for both 
channels. 
For the odd channel, $\omega_{\vQ}$ has a broad maximum around 
$\D=0.15$ and merges into the lower edge of particle-hole continuum; 
the particle-hole bound state is realized for $\D \lesssim 0.17$ 
(solid circles in \fig{d-wQ}). 
For the even channel the bound state is realized in the same doping 
region as for the odd channel (solid triangles), but appears inside 
the continuum in $0.12\lesssim \D\lesssim 0.17$ accompanied by the 
discontinuous change of $\omega_{\vQ}^{\rm res}$. 
This jump is due to the fact that the even channel consists of two 
different intraband scattering processes, within the bonding band 
and within the antibonding band.

\subsubsection{$\vq$ maps for $\alpha=0$} 

\begin{figure}
\centerline{\includegraphics[width=0.48\textwidth]{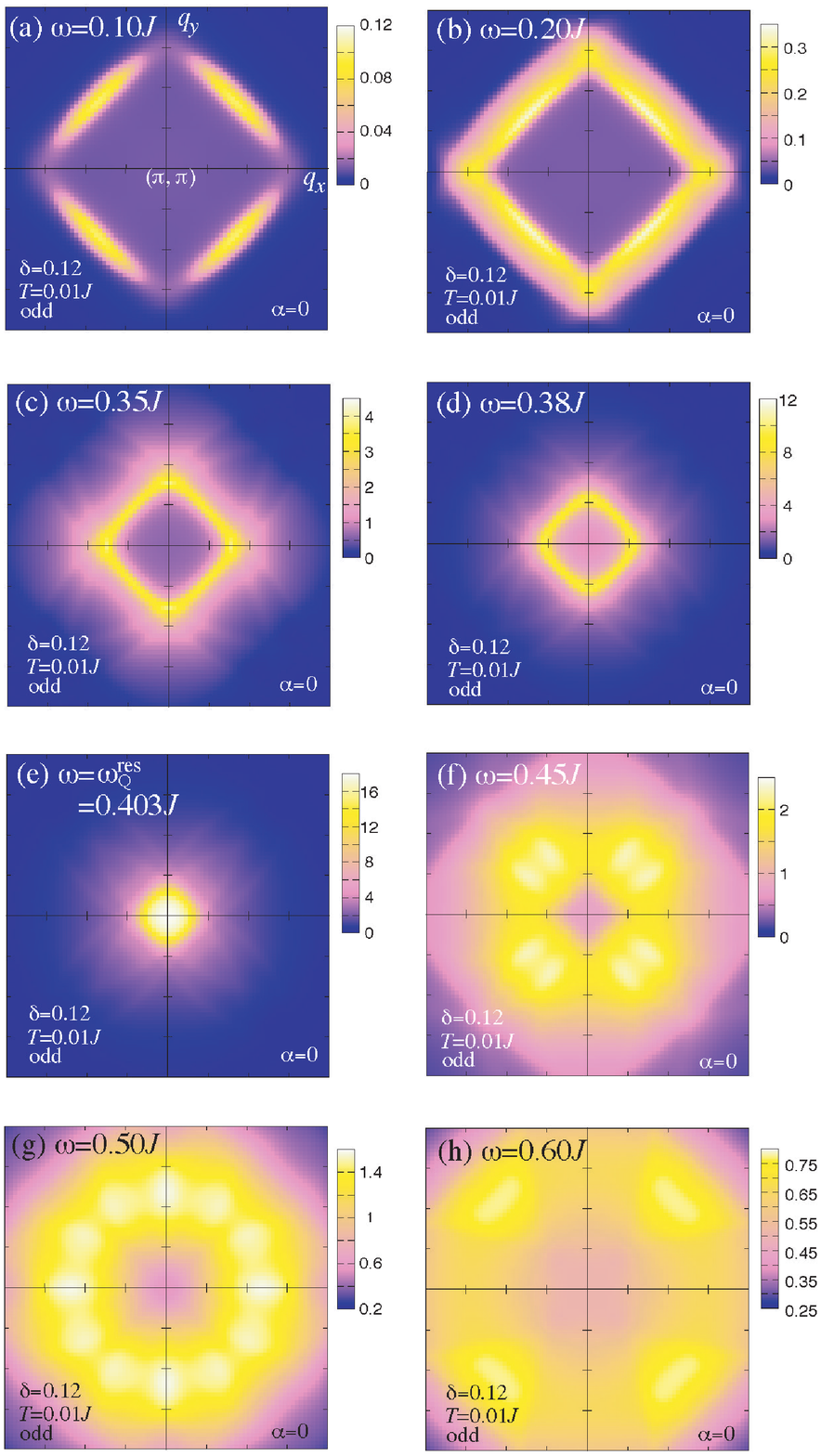}}
\caption{(color online) 
$\vq$ maps of Im$\chi(\vq,\,\omega)$ for a sequence of energies $\omega$ 
in $0.6\pi \leq q_{x}, q_{y} \leq 1.4\pi$ for  
$\D=0.12$, $T=0.01J$, and $\alpha=0$ in the odd channel. 
}
\label{qa0-w}
\end{figure}

We next show $\vq$ maps of Im$\chi(\vq,\,\omega)$ around $\vq=\vQ$ 
for a sequence of energies $\omega$ for $\D=0.12$.  
At low $\omega$  [\fig{qa0-w}(a)], strong spectral weight is localized 
in the region around $\vq=(\pi\pm2\pi\eta,\,\pi\pm2\pi\eta)$, 
which is due to the particle-hole excitations around the $d$-wave gap 
nodes. 
With increasing $\omega$, the scattering processes with 
$\vq=(\pi\pm2\pi\eta,\,\pi)$ and $(\pi,\,\pi\pm2\pi\eta)$ begin to 
contribute and the strong intensity region forms a diamond 
shape [\fig{qa0-w}(b)].   
This diamond shrinks when $\omega$ increases further [\fig{qa0-w}(c)], 
starts being due to collective excitations [\fig{qa0-w}(d)], and is 
finally reduced to the collective commensurate peak at 
$\omega=\omega_{\vQ}^{\rm res}=0.403J$ [\fig{qa0-w}(e)]. 
Above $\omega_{\vQ}^{\rm res}$ [\fig{qa0-w}(f)], the DIC spectral 
weight becomes dominant with rather broad features. 
While the spectral weight difference between the DIC and IC positions 
is not sizable and the DIC structures are smeared out in a certain 
energy range [\fig{qa0-w}(g)], the DIC signals seem to disperse 
outwards with increasing $\omega$ [Figs.~\ref{qa0-w}(f)-(h)] 
as expected from \fig{qwdetail}.

\begin{figure}
\centerline{\includegraphics[width=0.48\textwidth]{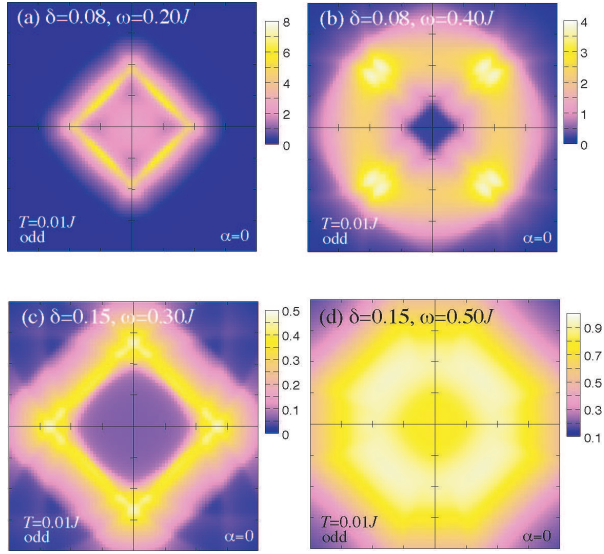}}
\caption{(color online) $\vq$ maps of Im$\chi(\vq,\,\omega)$ 
in $0.6\pi \leq q_{x}, q_{y} \leq 1.4\pi$ 
for $\D=0.08$ (upper panels) and $0.15$ (lower panels) for 
$T=0.01J$, and $\alpha=0$ in the odd channel. 
The energy is set below (above) $\omega_{\vQ}^{\rm res}$ 
in left (right) panels: 
$\omega_{\vQ}^{\rm res}=0.245J$ and $0.421J$ for 
$\D=0.08$ and $0.15$, respectively. 
}
\label{qa0-d}
\end{figure}

The spectral weight distribution in $\vq$-space for other doping 
rates, $\D=0.08$ and $0.15$, is shown in \fig{qa0-d} for 
$\omega < \omega_{\vQ}^{\rm res}$ (left panels) 
and $\omega > \omega_{\vQ}^{\rm res}$ (right panels). 
For $\omega<\omega_{\vQ}^{\rm res}$ 
we see that the strong weight forms a diamond shape independent of 
$\D$, and the IC signals become stronger 
than the DIC peaks for higher $\D$. 
For $\omega>\omega_{\vQ}^{\rm res}$, 
on the other hand, the DIC structure is clearly seen at relatively 
low $\D$, while the spectral weight spreads broader with $\D$ 
(see also \fig{qwa0-d}). 

\begin{figure}
\centerline{\includegraphics[width=0.95\textwidth]{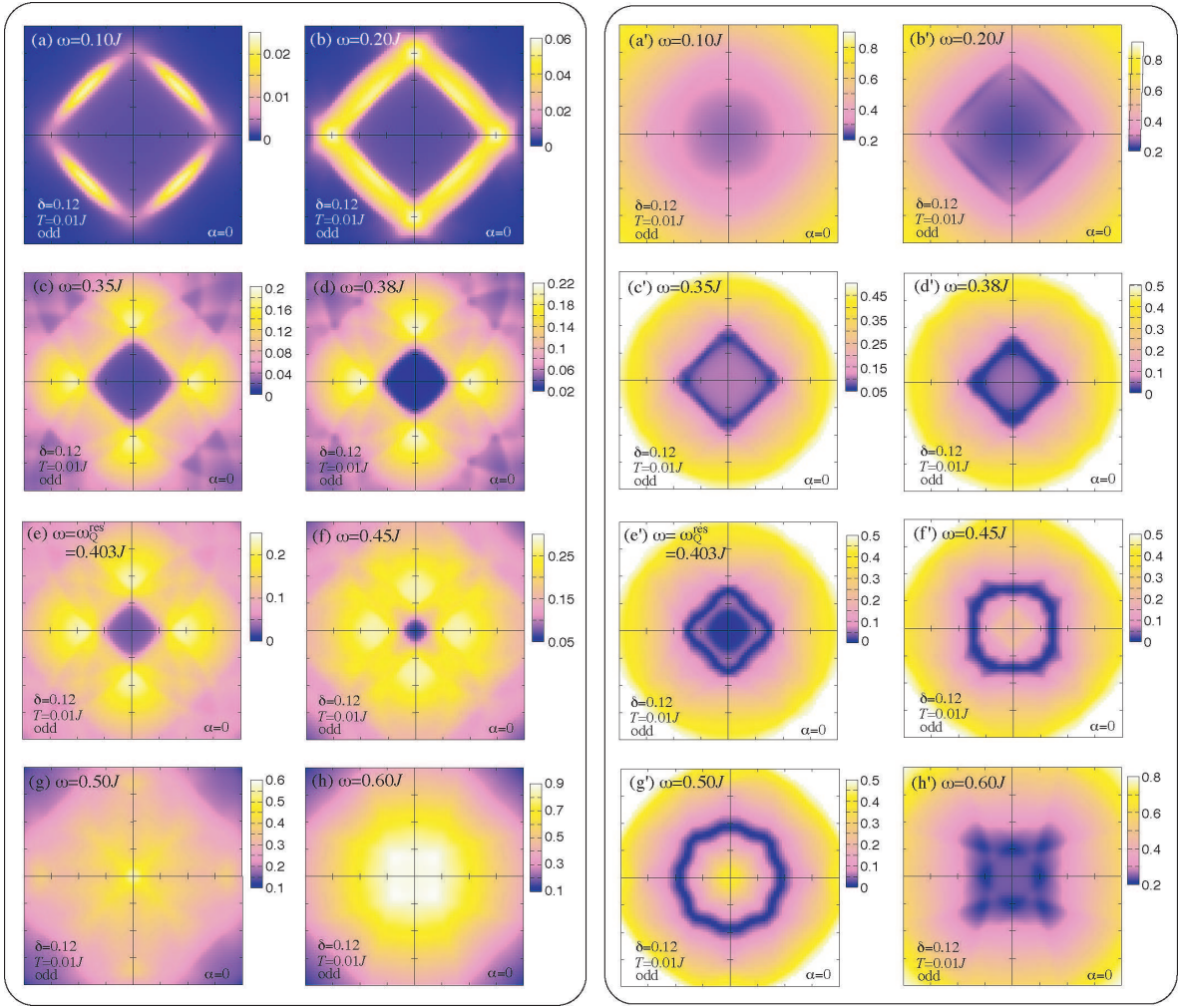}}
\caption{(color online) 
$\vq$ maps of $|J(\vq)$Im$\chi_{0}(\vq,\,\omega)|$ (left panel) 
and $|1+J(\vq)$Re$\chi_{0}(\vq,\,\omega)|$ (right panel) 
for a sequence of energies $\omega$ 
in $0.6\pi\leq q_{x}, q_{y} \leq 1.4\pi$ for   
$\D=0.12$, $T=0.01J$, and $\alpha=0$ in the odd channel. 
In (c')-(g'), the maps are restricted to $\vq$ with 
$|1+J(\vq)$Re$\chi_{0}(\vq,\,\omega)| \leq 0.5$ 
to get a better contrast in the interesting region. 
}
\label{qa0-wsup}
\end{figure}

To understand the intensity distribution in \fig{qa0-w}, we write 
the imaginary part of \eq{RPA} as
\be
{\rm Im}\chi(\vq,\,\omega)=\frac{{\rm Im}\chi_{0}(\vq,\,\omega)}
{[1+J(\vq){\rm Re}\chi_{0}(\vq,\,\omega)]^{2}+
[J(\vq){\rm Im}\chi_{0}(\vq,\,\omega)]^{2}}\,. \label{RPA2}
\ee
We have to consider both $|J(\vq){\rm Im}\chi_{0}(\vq,\,\omega)|$ and 
$|1+J(\vq){\rm Re}\chi_{0}(\vq,\,\omega)|$, which we show in the 
left panel and the right panel in \fig{qa0-wsup}, respectively, 
at the same $\omega$ as in \fig{qa0-w};  
the structure of Im$\chi_{0}(\vq,\,\omega)$ is essentially the same 
as that of $J(\vq){\rm Im}\chi_{0}(\vq,\,\omega)$. 
Two typical energy scales, 
$\omega_{\vQ}^{\rm res}$ and $\omega_{\rm th}(\vQ)$, characterize 
the spectral weight distribution of Im$\chi(\vq,\,\omega)$. 

(1) For $\omega\ll \omega_{\vQ}^{\rm res}$, we find 
$|1+J(\vq){\rm Re}\chi_{0}(\vq,\,\omega)| \gg 
 |J(\vq){\rm Im}\chi_{0}(\vq,\,\omega)|$
in Figs.~\ref{qa0-wsup}(a), (a'), (b), and (b'). 
Since the $\vq$ dependence of Re$\chi_{0}(\vq,\,\omega)$ is weak around 
$\vq=\vQ$ we have 
${\rm Im}\chi(\vq,\,\omega) \propto {\rm Im}\chi_{0}(\vq,\,\omega)$, 
that is, the weight distribution in Figs.~\ref{qa0-w}(a) and (b) 
is determined mainly by Im$\chi_{0}(\vq,\,\omega)$. 

(2) For $\omega\lesssim \omega_{\vQ}^{\rm res}$ 
[Figs.~\ref{qa0-wsup}(c) and (c')], the minimum values of 
$|1+J(\vq){\rm Re}\chi_{0}(\vq,\,\omega)|$, become comparable 
with $|J(\vq){\rm Im}\chi_{0}(\vq,\,\omega)|$ in a certain $\vq$ region. 
From \eq{RPA2}, therefore, a strong weight of Im$\chi(\vq\,\,\omega)$ 
is determined by the minimum position of 
$|1+J(\vq){\rm Re}\chi_{0}(\vq,\,\omega)|$, not by peak position of 
Im$\chi_{0}(\vq,\,\omega)$. 
Since Im$\chi_{0}(\vq,\,\omega)$ does not have an apparent structure 
in the region where $|1+J(\vq){\rm Re}\chi_{0}(\vq,\,\omega)|$ takes  
minima, it depends on details which signal can be stronger, 
the IC or the DIC. 
In our model, the IC signals tend to be stronger than the DIC 
signals at higher $\D$. 

(3) At $\omega \approx \omega_{\vQ}^{\rm res}$, 
$|1+J(\vq){\rm Re}\chi_{0}(\vq,\,\omega)|$ can vanish for suitable $\vq$ 
vectors [Figs.~\ref{qa0-wsup}(d') and (e')], leading to 
Im$\chi(\vq,\,\omega) \propto {\rm Im}\chi_{0}(\vq,\,\omega)^{-1}$. 
Since Im$\chi_{0}(\vq,\,\omega)$ also can become very small 
in the same $\vq$ region [Figs.~\ref{qa0-wsup}(d) and (e)], 
the strongest weight of Im$\chi(\vq,\,\omega)$ appears there, 
corresponding to a collective mode.  

(4) For $\omega_{\vQ}^{\rm res} \lesssim \omega \lesssim 
\omega_{\rm th}(\vQ)$, $|1+J(\vq){\rm Re}\chi_{0}(\vq,\,\omega)|$ 
vanishes in the region where Im$\chi_{0}(\vq,\,\omega)$ has 
IC peaks [Figs.~\ref{qa0-wsup}(f) and (f')]. 
Since Im$\chi(\vq,\,\omega) \propto {\rm Im}\chi_{0}(\vq\,,\omega)^{-1}$ 
when $1+J(\vq){\rm Re}\chi_{0}(\vq,\,\omega) = 0$,
the IC spectral weight is substantially diminished and 
Im$\chi(\vq,\,\omega)$ shows DIC peaks as seen in \fig{qa0-w}(f). 

(5) At $\omega \approx \omega_{\rm th}(\vQ)$, 
the IC peaks of $|J(\vq){\rm Im}\chi_{0}(\vq,\,\omega)|$ 
merge into a commensurate peak as shown in \fig{qa0-wsup}(g). 
$|J(\vq){\rm Im}\chi_{0}(\vq,\,\omega)|$ 
does not have an apparent structure around the region where 
$|1+J(\vq){\rm Re}\chi_{0}(\vq,\,\omega)|$ vanishes [\fig{qa0-wsup}(g')].  
Hence the strong weight of Im$\chi(\vq\,,\omega)$ has a similar 
distribution to that of $1+J(\vq){\rm Re}\chi_{0}(\vq,\,\omega)\approx 0$, 
and shows a ``ring'' in \fig{qa0-w}(g).

(6) For $\omega\gtrsim\omega_{\rm th}(\vQ)$, Im$\chi(\vq,\,\omega)$ has 
strongest intensity in the region where
$|1+J(\vq){\rm Re}\chi_{0}(\vq,\,\omega)| \approx 
 |J(\vq){\rm Im}\chi_{0}(\vq,\,\omega)|$  
[compare \fig{qa0-w}(h) with Figs.~\ref{qa0-wsup}(h) and (h')]. 

The spectral distribution of Figs.~\ref{qa0-d}(a)-(d) 
corresponds to the cases (2), (4), (2), and (6), respectively.

\subsubsection{$T$ dependence for $\alpha=0$} 

\begin{figure}
\centerline{\includegraphics[width=0.45\textwidth]{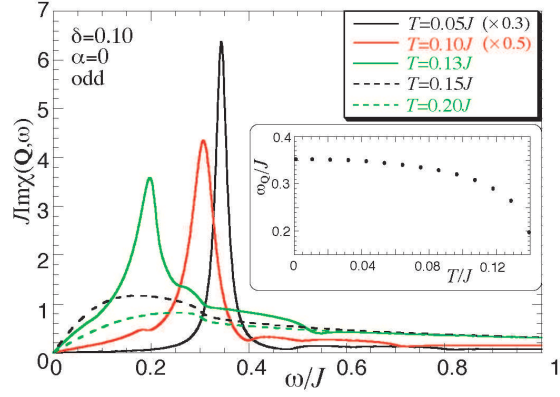}}
\caption{(color online) 
$\omega$ dependence of Im$\chi(\vQ,\,\omega)$ for several 
choices of $T$ for $\D=0.10$, and $\alpha=0$ in the odd channel; 
here $T_{\rm RVB}=0.139J$;   
for $T=0.05J$ and $0.10J$, Im$\chi(\vQ,\,\omega)$ is multiplied by 
0.3 and 0.5, respectively. 
The inset shows the $T$ dependence of the peak energy of 
Im$\chi(\vQ,\,\omega)$. 
}
\label{wTscan}
\end{figure}

So far we have presented results for a fixed temperature $T=0.01J$. 
Now we discuss the $T$ dependence of Im$\chi(\vq,\,\omega)$. 
In \fig{wTscan}, we plot Im$\chi(\vq,\,\omega)$ at $\vq=\vQ$ 
as a function of $\omega$ for several choices of $T$ for $\D=0.10$
in the odd channel. 
The $d$-wave gap disappears at $T_{\rm RVB}= 0.139J$ for this 
doping.  
We see that the sharp peak of Im$\chi(\vQ,\,\omega)$ survives 
as long as the $d$-wave gap exists. 
The peak energy is determined mainly by minimizing 
$|1+J(\vQ){\rm Re}\chi_{0}(\vQ,\,\omega)|$. 
The $T$ dependence of the peak energy is shown in the inset of 
\fig{wTscan}; it follows approximately the $T$ dependence of 
the $d$-wave gap [see \fig{meana0}(a)]. 
With increasing $T$ low-energy spectral weight proportional to 
$\omega$ becomes noticeable, and Im$\chi(\vQ,\,\omega)$
evolves smoothly toward the spectrum in the normal state 
($T>T_{\rm RVB}$), where the sharp peak at finite $\omega$
disappears. 

\begin{figure}
\centerline{\includegraphics[width=0.48\textwidth]{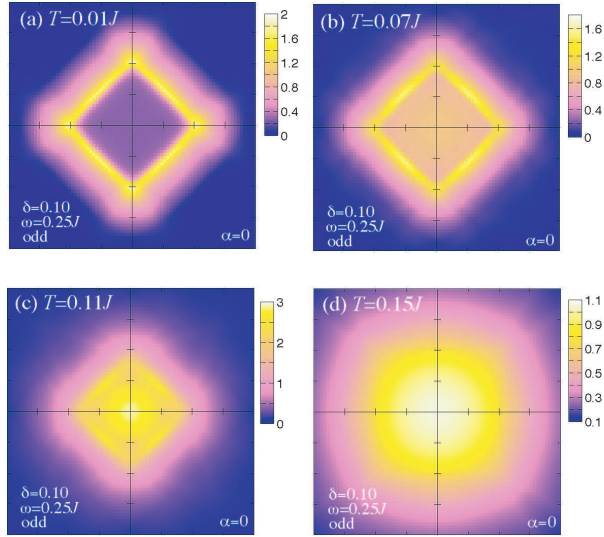}}
\caption{(color online) 
$\vq$ maps of Im$\chi(\vq,\,\omega)$ for a sequence of temperatures 
in $0.6\pi \leq q_{x}, q_{y} \leq 1.4\pi$ 
for $\D=0.10$, $\omega=0.25J$, and $\alpha=0$ in the odd channel; 
here $T_{\rm RVB}=0.139J$. 
}
\label{qa0-T}
\end{figure}

The temperature dependence of $\vq$ maps is shown in \fig{qa0-T} 
for several choices of $T$ at $\D=0.10$ and $\omega=0.25J$. 
Strong intensity of Im$\chi(\vq,\,\omega)$ forms a diamond shape 
at low $T$. 
Although the diamond shape is retained with increasing $T$, 
the weight inside the diamond becomes stronger, leading to a dominant 
commensurate signal above $T\sim 0.11J$.  
The commensurate signal thus dominates already at tempertures
slightly below the onset temperature of $d$-wave singlet pairing 
$T_{\rm RVB}= 0.139J$.

\begin{figure}
\centerline{\includegraphics[width=0.5\textwidth]{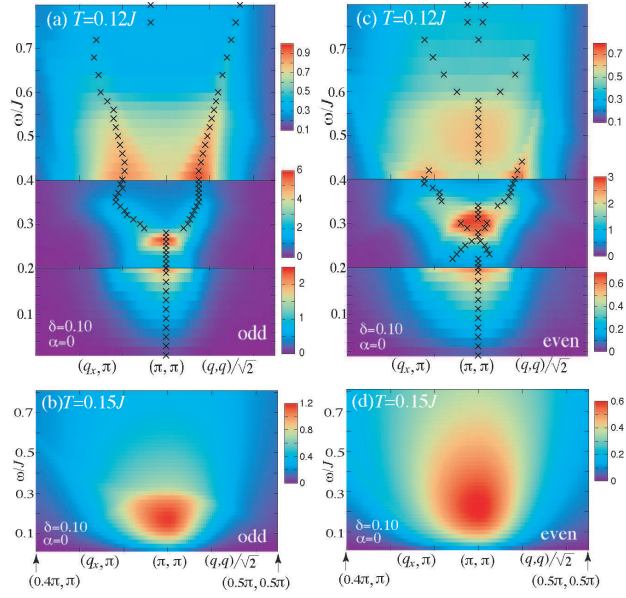}}
\caption{(color) 
$(\vq,\,\omega)$ maps of Im$\chi(\vq,\,\omega)$ at  
$T< T_{\rm RVB}=0.139J$ (a) and (c), and 
$T> T_{\rm RVB}$ (b) and (d) for $\D=0.10$ and $\alpha=0$; 
the left (right) panels are for the odd (even) channel;
$\vq$-scans as in \fig{qwa0-d}.
The cross symbols in (a) and (c) 
represent the highest weight positions 
along $\vq=(q_{x},\,\pi)$ and $\frac{1}{\sqrt{2}}(q,\,q)$.  
}
\label{uqwa0}
\end{figure}

A $(\vq,\,\omega)$ map of ${\rm Im}\chi(\vq,\,\omega)$ is shown in 
\fig{uqwa0}(a) for $T=0.12J$ ($<T_{\rm RVB}$), where the energy region 
is separated into three with different color scales; the cross symbols 
mark highest weight positions along $\vq=(q_{x},\,\pi)$ and 
$\frac{1}{\sqrt{2}}(q,\,q)$. 
The strongest weight appears at $\vq=\vQ$ and $\omega\approx 0.26J$. 
Below this energy IC structures are not realized any longer and 
the spectrum shows a broad commensurate structure. 
However, in the high energy region ($\omega \gtrsim 0.30J$) DIC 
signals still survive as well as IC signals. 
When $T$ is further increased above $T_{\rm RVB}$ [\fig{uqwa0}(b)], 
neither IC nor DIC structures are realized. 
The strong intensity is centered around $\vq=\vQ$ and 
broadly spreads out. 
The right panels in \fig{uqwa0} are corresponding results 
for the even channel. 
Overall features look similar as for the odd channel except that 
the spectral weight becomes rather broadened. 
Although  the peak positions form a rather complex pattern in 
\fig{uqwa0}(c), an upward dispersion for  
$0.3J \lesssim \omega \lesssim 0.45J$ can still be recognized 
together with a remnant of the downward dispersion for 
$0.2J \lesssim \omega \lesssim 0.25J$.

\subsubsection{$\omega$ dependence of $\vq$ maps for $\alpha\neq 0$}
 
In the following three subsections,
we present results for Im$\chi(\vq,\omega)$ for $\alpha=-0.05$ to
discuss effects due to an in-plane anisotropy and their enhancement
by $d$FSD correlations.

\begin{figure}
\centerline{\includegraphics[width=0.48\textwidth]{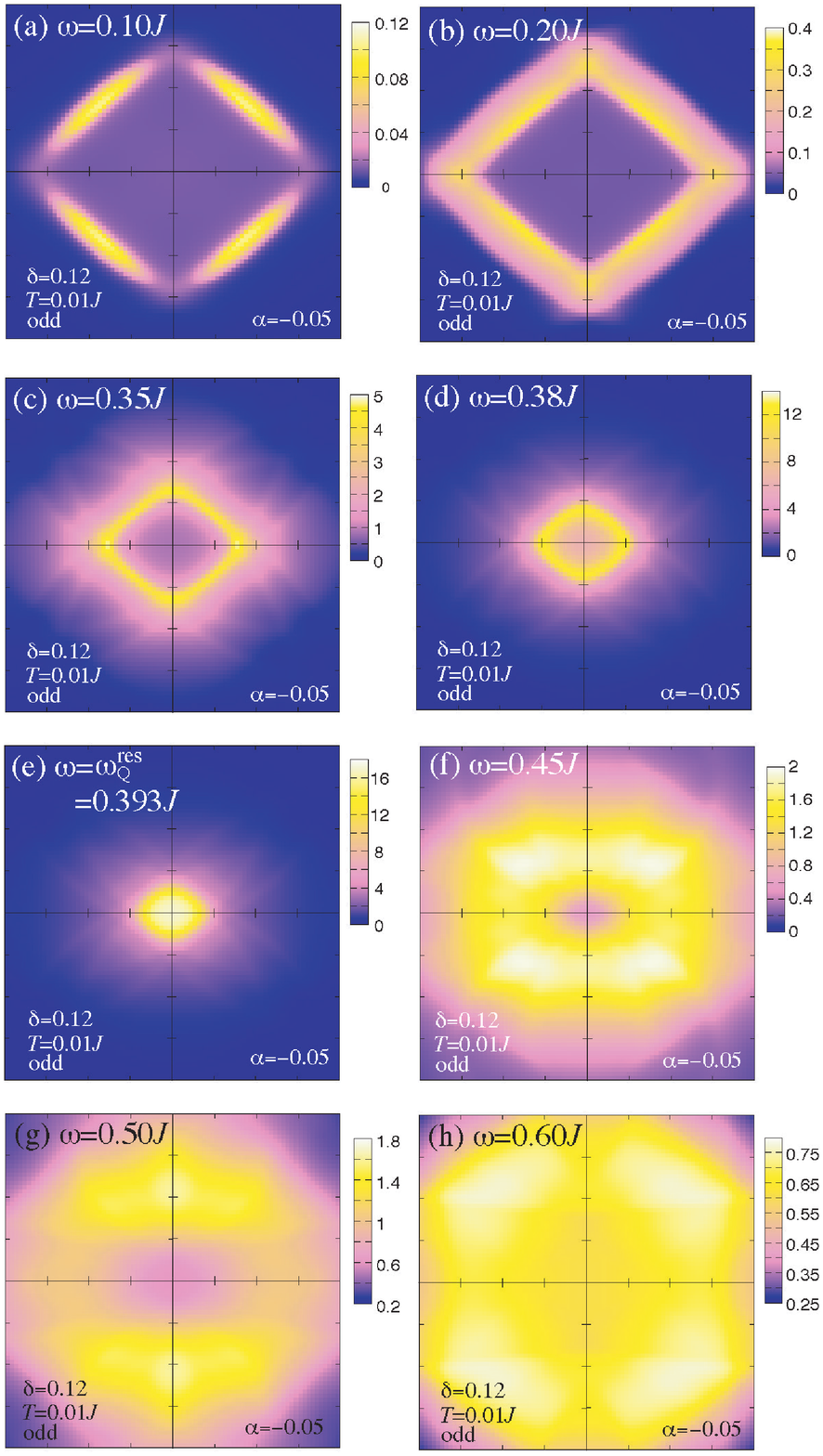}}
\caption{(color online) 
$\vq$ maps of Im$\chi(\vq,\,\omega)$ for a sequence of $\omega$ 
in $0.6\pi \leq q_{x}, q_{y} \leq 1.4\pi$ for 
$\D=0.12$, $T=0.01J$, and $\alpha=-0.05$ in the odd channel. 
}
\label{qa5-w}
\end{figure}

Figure~\ref{qa5-w} shows $\vq$ maps of Im$\chi(\vq,\,\omega)$ 
for a sequence of energies $\omega$ for $\D=0.12$ and $T=0.01J$; 
the corresponding results for $\alpha=0$ were shown in \fig{qa0-w}. 
At low $\omega$, the DIC signals appear, and form a high intensity 
region which is a bit distorted compared to \fig{qa0-w}(a), due to 
the in-plane anisotropy. 
These DIC signals come from nodal scatterings. 
Note that the nodes of the singlet pairing deviate slightly 
from the diagonal direction because of $\Delta_{x} \neq \Delta_{y}$ 
and finite $\Delta_{z}$ in Eqs.~(\ref{insc}) and (\ref{outsc}). 
Figures~\ref{qa5-w}(b)-(d) are results for larger
$\omega<\omega_{\vQ}^{\rm res}$. 
Appreciable weight appears on both the $q_{x}$ and $q_{y}$ axes and 
in this sense a nearly two-dimensional excitation spectrum is obtained 
even if the antibonding FS becomes open as in \fig{FS}.
Analogous behavior was already found for the single layer $t$-$J$ model 
in Ref.~\cite{yamase01}. 
Effects of the anisotropy are seen in (i) the difference of the IC peak 
positions, namely $(\pi\pm2\pi\eta_{x},\,\pi)$ and 
$(\pi,\,\pi\pm2\pi\eta_{y})$ with $\eta_{x}\neq \eta_{y}$, 
leading to a deformed diamond shape, 
and (ii) the relative peak intensity difference of the two IC peaks. 
The latter however depends strongly on  $\omega$ and $\D$ even if 
$\alpha$ is fixed.  
For example, the IC peak at $\vq=(\pi,\,\pi\pm2\pi\eta_{y})$ is a bit 
higher than that at $\vq=(\pi\pm2\pi\eta_{x},\,\pi)$ at $\omega=0.20J$ 
[\fig{qa5-w}(b)], while the order is reversed at $\omega=0.35J$ 
[\fig{qa5-w}(c)]. 
Moreover, as we shall discuss in Sec.V.B., the peak intensity difference 
can depend strongly on details of the band structures and thus on models. 
On the other hand, the former is a robust feature of 
Im$\chi(\vq,\,\omega)$. When the difference of the IC peak position, 
$\Delta\eta = \eta_{x}-\eta_{y}$, is taken as a measure of the anisotropy 
of the IC peaks, we have $\Delta\eta > 0$ for $\alpha<0$ and 
in addition $\Delta\eta \propto \alpha$ in good approximation 
at least up to $|\alpha| \sim 0.1$. 
Furthermore, the $\omega$ dependence of $\Delta\eta$ is weak. 
In \fig{eta-w} we plot $\eta_{x}$ and $\eta_{y}$ as a function of 
$\omega$. While $\eta_{x}$ and $\eta_{y}$ depend on $\omega$,  
one finds $\Delta\eta > 0$ for all $\omega < \omega_{\vQ}^{\rm res}$, 
and a weak $\omega$ dependence of $\Delta\eta$ except for the region 
of $\omega \approx \omega_{\vQ}^{\rm res}$ where both $\eta_x$ and
$\eta_y$ tend to zero. 

Figure~\ref{qa5-w}(e) corresponds to the energy 
$\omega=\omega_{\vQ}^{\rm res}$. 
While the strongest weight appears at $\vq=\vQ$ for that energy, 
Im$\chi(\vq,\omega_{\vQ}^{\rm res})$ has an apparent elliptic 
distribution in the $\vq$ plane. 
Figures~\ref{qa5-w}(f)-(h) are for $\omega>\omega_{\vQ}^{\rm res}$.  
In \fig{qa5-w}(f), the strongest weight appears along the diagonal 
direction, which is a remnant of the DIC signal 
for $\alpha=0$ [\fig{qa0-w}]. 
The strong weight distribution, however, forms a rectangular shape 
and shows a pronounced anisotropy. 
As energy is increased a little, Im$\chi(\vq,\,\omega)$ shows 
one-dimensional like signatures in \fig{qa5-w}(g) even with a small 
anisotropy (5\%). 
To understand this one-dimensional pattern, we go back to 
Figs.~\ref{qa0-wsup} (g) and (g'). 
For $\alpha=-0.05$, Im$\chi_{0}(\vq,\,\omega)$ still shows a commensurate 
signal but the spectral weight spreads out largely along the $q_{x}$ 
direction. Since $|1+J(\vq)$Re$\chi_{0}(\vq,\,\omega)|$ vanishes around 
$\vq=\vQ$ as shown in \fig{qa0-wsup}(g'), the resulting 
Im$\chi(\vq,\,\omega)$ shows enhanced spectral weight along 
the $q_{y}$ direction, leading to \fig{qa5-w}(g). 
With further increasing $\omega$, however, the spectral weight becomes 
much broader [\fig{qa5-w}(h)] and the anisotropy becomes less apparent.  
For $\omega > \omega_{\vQ}^{\rm res}$ therefore the spectral weight 
distribution forms various patterns and is very sensitive to energy. 
This is in sharp contrast to the robust feature of the deformed diamond 
shape distribution for $\omega < \omega_{\vQ}^{\rm res}$, which enables 
us to define $\Delta\eta$ as a measure of the anisotropy of the magnetic 
excitations.

\subsubsection{$\D$ dependence of $\vq$ maps for $\alpha\neq 0$}

\begin{figure}
\centerline{\includegraphics[width=0.48\textwidth]{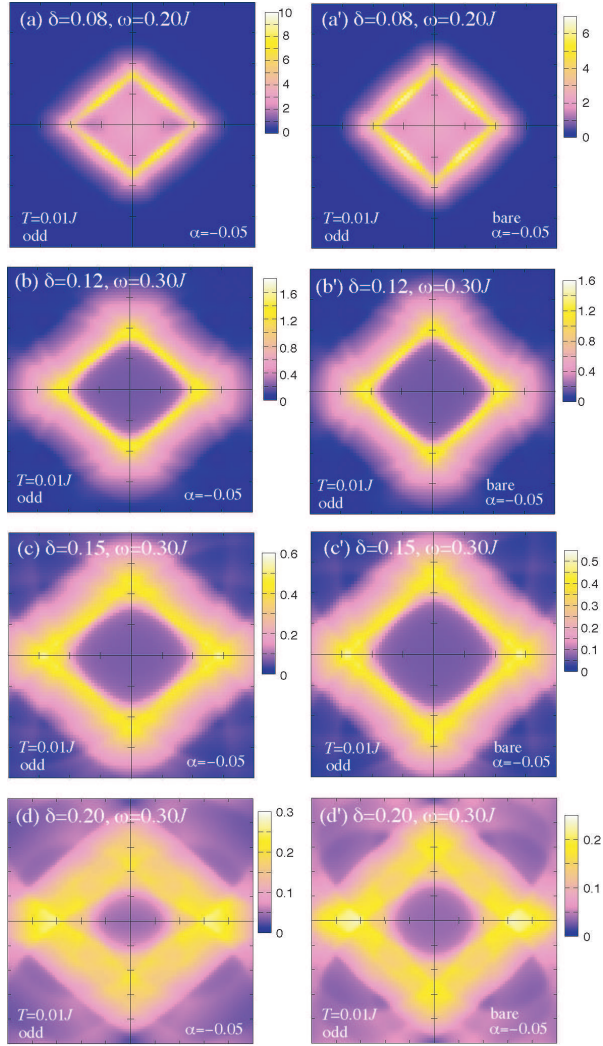}}
\caption{(color online) 
$\vq$ maps of Im$\chi(\vq,\,\omega)$ for a sequence of doping 
concentrations $\D$ for $T=0.01J$, and $\alpha=-0.05$ in the odd channel; 
$\omega$ is chosen below $\omega_{\vQ}^{\rm res}$; 
the right panels show the results obtained from the bare anisotropy 
without the enhancement due to $d$FSD correlations; 
$\vq$ is  scanned in $0.6\pi \leq q_{x}, q_{y} \leq 1.4\pi$ 
except for the panels for $\D=0.20$ where 
$0.5\pi\leq q_{x}, q_{y} \leq 1.5\pi$ is taken. 
}
\label{qa5-d}
\end{figure}

The left panels of \fig{qa5-d} show $\vq$ maps of Im$\chi(\vq,\,\omega)$ 
for a sequence of doping concentrations $\D$; the energy $\omega$ 
is below $\omega_{\vQ}^{\rm res}$ (actually below $\omega_{\vQ}$ for
$\D = 0.20$, where no collective mode exists, see \fig{d-wQ}).
Appreciable intensity appears along a deformed diamond for all $\D$. 
Although the renormalized band anisotropy decreases with increasing $\D$ 
[\fig{meana5}(b)], the anisotropy of Im$\chi(\vq,\,\omega)$ is
enhanced. 
To identify the influence of $d$FSD correlations, we compare 
these results with the corresponding results for the bare anisotropy
on the right panels of \fig{qa5-d}.  
The latter are calculated by switching off the $d$FSD correlations, 
that is, we impose the same bare anisotropy $\alpha=-0.05$, but use 
mean fields for $\alpha=0$.
We see that the effect of the bare anisotropy on Im$\chi(\vq,\,\omega)$ 
is doping dependent even for fixed (doping-independent) $\alpha$ and 
becomes less pronounced at lower $\D$. 
In particular, the diamond is almost symmetric for $\D=0.08$ and $0.12$ 
[Figs.~\ref{qa5-d}(a') and (b')]. 
That is, the $d$FSD correlations are particularly important to 
understand the anisotropy of Im$\chi(\vq,\,\omega)$ at lower $\D$ 
[Figs.~\ref{qa5-d}(a) and (b)]. 
This holds for both odd and even channels and similar results are 
obtained for the even channel.

\subsubsection{$T$ dependence of $\vq$ maps for $\alpha\neq 0$}

\begin{figure}
\centerline{\includegraphics[width=0.95\textwidth]{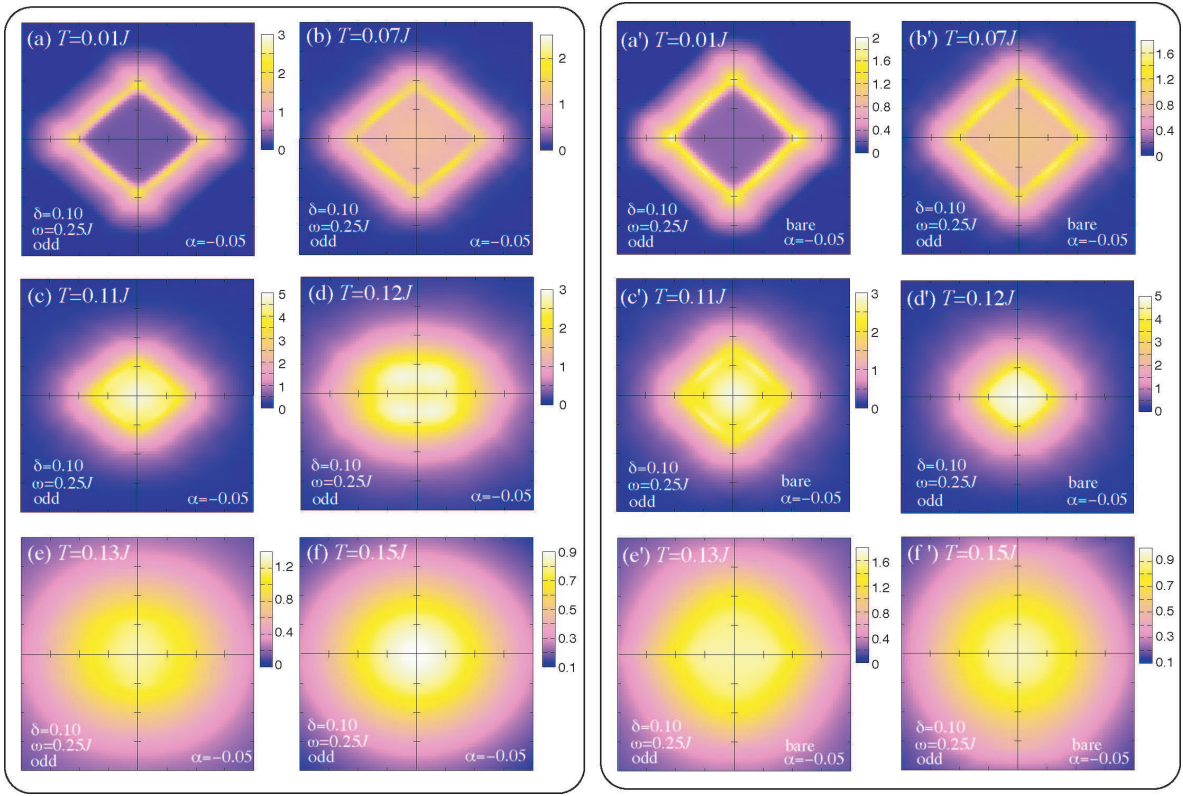}}
\caption{(color online) 
Left panel:  
$\vq$ maps of Im$\chi(\vq,\,\omega)$ for a sequence of $T$ 
in $0.6\pi\leq q_{x}, q_{y} \leq 1.4\pi$ for   
$\D=0.10$, $\omega=0.25J$, and $\alpha=-0.05$
in the odd channel; here $T_{\rm RVB}=0.133J$.  
Right panel: corresponding results for the bare anisotropy, that is,  
without $d$FSD correlations; $T_{\rm RVB}=0.139J$ in that case. 
}
\label{qa5-T}
\end{figure}

Momentum space maps of Im$\chi(\vq,\,\omega)$ for a sequence of 
temperatures are shown in the left panel of \fig{qa5-T} for $\D=0.10$ 
and $\omega=0.25J$. 
The deformed diamond shaped distribution is rather robust against 
$T$ [Figs.~\ref{qa5-T}(a) and (b)], although the spectral weight 
inside the diamond gradually increases with $T$,
as we have seen in \fig{qa0-T}. 
At temperatures slightly below $T_{\rm RVB}=0.133J$, the commensurate 
signal becomes dominant [\fig{qa5-T}(c)] and a pronounced anisotropy 
appears in Fig.~\ref{qa5-T}(d). 
When $T$ increases further, the anisotropy is however reduced
[Figs.~\ref{qa5-T}(e) and (f)].  
The temperature dependence of the anisotropy in Fig.~\ref{qa5-T}
is not monotonically linked to the $T$-dependence of the band 
anisotropy. 
As seen in \fig{meana5}(b) for $\D=0.10$, the band anisotropy 
at $T=0.13J$ is larger than that at $T=0.12J$, but does not lead 
to a stronger anisotropy of Im$\chi(\vq,\,\omega)$ at $T=0.13J$. 

To analyze the relevance of $d$FSD correlations, we calculate 
Im$\chi(\vq,\,\omega)$ also for isotropic mean fields and show
the results in the right panel of \fig{qa5-T}. 
We see that while $d$FSD correlations contribute to the enhancement 
of the anisotropy of Im$\chi(\vq,\,\omega)$ at any $T$, such an 
effect becomes most pronounced at relatively high $T$  
as seen in Figs.~\ref{qa5-T}(d) and (d'). 
While \fig{qa5-T} has been obtained for fixed $\omega$ and $\D$,
results for other parameter sets including the even channel 
show that the $d$FSD correlations
typically drive a pronounced anisotropy at relatively 
high $T$ ($\sim T_{\rm RVB}$).

\section{Discussion}

We have investigated Im$\chi(\vq,\,\omega)$ in the slave-boson 
mean-field scheme for the bilayer $t$-$J$ model. 
We summarize our results through a comparison with inelastic 
neutron scattering data for YBCO. 
Experimental data are well summarized in 
Ref.~\cite{rossatmignod91,bourges98,fong00,dai01}. 
It should be kept in mind that in the slave-boson mean-field 
theory\cite{fukuyama98} $T_{\rm RVB}$ has to be interpreted as  
pseudogap crossover temperature $T^{*}$ in the underdoped regime, 
and as superconducting phase transition temperature $T_{c}$ 
in the overdoped regime of high-$T_{c}$ cuprates.

\subsection{Prominent features of magnetic excitations in YBCO} 

One of the most prominent features of magnetic excitations 
in YBCO is that IC peaks appear only below $T_{c}$ or 
$T^{*}$,\cite{dai98,arai99,bourges00,dai01,stock04} 
which is in sharp contrast to LSCO where IC peaks were reported 
even at room temperature.\cite{thurston89,aeppli97} 
This is captured by the present theory.  
As shown in \fig{qa0-T}, the IC signals are obtained only at 
low $T$ and change into a broad commensurate signal above 
$T \sim T_{\rm RVB}$. 
It was pointed out a few years ago that the distinct behavior 
of LSCO is also captured by the present theoretical framework 
with different band parameters.\cite{yamase00,yamase01} 

\begin{figure}

\centerline{\includegraphics[width=0.48\textwidth]{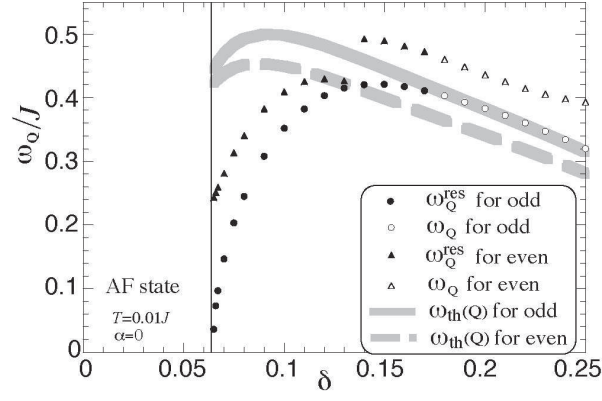}}
\caption{
Doping dependence of $\omega_{\vQ}$ for the odd and even 
channels at $T=0.01J$ and $\alpha=0$; 
the superscript ``res'' indicates that \eq{wq1} is satisfied 
at $\vq=\vQ$ and $\omega=\omega_{\vQ}$;
$\omega_{\rm th}(\vQ)$ is also plotted with bold gray lines for 
both channels. 
}
\label{d-wQ}
\end{figure}

Another central issue in YBCO is the so-called resonance peak, 
which was reported as a sharp magnetic signal at $\vq=\vQ$ 
and finite $\omega$.\cite{rossatmignod91,mook93,bourges95,
fong9596,bourges96,dai96,fong97}  
There are several alternative scenarios for this resonance 
peak.\cite{lavagna94,blumberg95,demler98,batista01,vojta06} 
In the present theory, in accordance with previous 
work,\cite{liu95,bulut96,millis96,manske0198,brinckmann99,
norman00,kao00,brinckmann01,
chubukov01,tchernyshyov01,li02,onufrieva02,sega0306,schnyder04,
zhou0405,eremin05,schnyder06} 
the resonance is interpreted as a particle-hole bound state   
(Figs.~\ref{wthfig} and \ref{qwa0-d}). 
While the resonance is particularly pronounced at low $T$
($\ll T_{\rm RVB}$), it starts to develop below $T_{\rm RVB}$ 
(Fig.~\ref{wTscan}), that is, below $T^{*}$ in the underdoped 
regime and below $T_{c}$ in the overdoped regime. 
This is consistent with experiments for optimally doped 
YBCO\cite{mook93,fong9596,bourges95,bourges96,bourges00} and 
YBa$_{2}$Cu$_{3}$O$_{6.6}$.\cite{dai99} 
The $\D$ dependence of $\omega_{\vQ}^{\rm res}$ (\fig{d-wQ})   
roughly agrees with experiments for both 
odd\cite{fong00,dai01,stock04,pailhes06} 
and even\cite{pailhes03,pailhes04,pailhes06} channels 
if we set $J \approx 100$meV.\cite{bourges98} 
Although this agreement is based on tuning one of the band 
parameters, namely $t/J$, it is remarkable that the $\D$ 
dependence of $\omega_{\vQ}^{\rm res}$ is reproduced within 
the present simple framework. 
We note that no genuine resonance was obtained for 
$\D \gtrsim 0.18$ in \fig{d-wQ}, which however depends on the 
choice of $\Gamma$. 

\begin{figure}
\centerline{\includegraphics[width=0.48\textwidth]{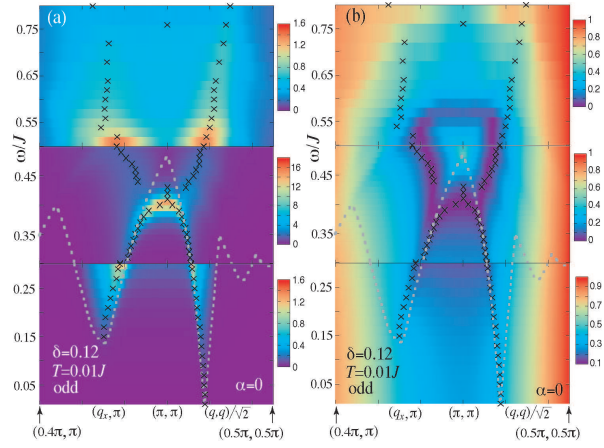}}
\caption{(color) 
$(\vq,\,\omega)$ maps of Im$\chi(\vq,\,\omega)$ (a) and 
$|1+J(\vq)$Re$\chi_{0}(\vq,\,\omega)|$ (b) 
for $\D=0.12$, $T=0.01J$, and $\alpha=0$ in the odd channel;
$\vq$-scans as in \fig{qwa0-d}.
The cross symbols represent the highest weight positions 
of Im$\chi(\vq,\,\omega)$ along $\vq=(q_{x},\,\pi)$ and 
$\frac{1}{\sqrt{2}}(q,\,q)$, respectively.
}
\label{qwdetail}
\end{figure}

As seen from Figs.~\ref{qwa0-d} and \ref{qwdetail}, the collective 
mode usually has a downward dispersion and is necessarily 
accompanied by IC signals for $\omega<\omega_{\vQ}^{\rm res}$. 
This downward dispersion was actually 
observed.\cite{arai99,bourges00,reznik04,pailhes04} 
The resonance at $\vq=\vQ$ and the IC peaks observed in experiments 
are therefore understood as coming from the same origin. 
The resonance mode, however, is realized only in a limited 
energy region in the presence of finite damping and temperature, 
and loses its collective nature when it touches the continuum 
spectrum (\fig{qwa0-d}). 

The IC peaks are well defined along the cut with $\vq=(q_{x},\,\pi)$ 
or $(\pi,\,q_{y})$ for $\omega \lesssim \omega_{\vQ}^{\rm res}$. 
In the $\vq$ plane, however, strong intensity weight appears on 
a diamond shaped region as we have seen in Figs.~\ref{qa0-w}(b)-(d). 
Hence, the DIC peaks are also well defined along a cut with 
$\vq=(q,\,q)/\sqrt{2}$. Although the possibility of DIC peaks 
was not discussed in the experimental literature, this diamond 
shaped distribution was actually observed.\cite{mook98}  
For $\omega \ll \omega_{\vQ}^{\rm res}$ the DIC signals become 
dominant [\fig{qa0-w}(a)], 
which is a robust feature coming from the $d$-wave gap nodes.  
These DIC signals were already predicted about a decade 
ago,\cite{zha93,tanamoto94} but have not been detected in 
experiments. 

For $\omega>\omega_{\vQ}^{\rm res}$, we have obtained 
an upward dispersion in Figs.~\ref{qwa0-d} and \ref{qwdetail} 
especially at relatively low $\D$ and have interpreted it 
as an overdamped collective mode; 
note that this mode is not directly connected with the downward  
collective mode in $\omega <\omega_{\vQ}^{\rm res}$.
An upward dispersion of DIC peaks for
$\omega > \omega_{\vQ}^{\rm res}$ can be read off also from 
Figs.~\ref{qa0-w} and \ref{qa0-d}. 
These DIC peaks are robust at least for
$\omega_{\vQ}^{\rm res} \lesssim \omega \lesssim \omega_{\rm th}(\vQ)$ 
as discussed in Figs.~\ref{qa0-wsup}(f) and (f'), 
although they seem to continue up to higher $\omega$ than 
$\omega_{\rm th}(\vQ)$ in Figs.~\ref{qwa0-d} and \ref{qwdetail}. 
An upward DIC dispersion was recently reported in neutron scattering 
experiments.\cite{reznik04,pailhes04,hayden04}

As seen in Figs.~\ref{uqwa0}(a) and (c), the upward dispersion 
in the high energy region is rather robust to temperature while 
in the low energy region temperature spoils the incommensurate 
structures in favor of a broad commensurate signal. 
This behavior was detected in the most recent neutron scattering 
experiments.\cite{hinkov06}  

The even channel behaves qualitatively similarly to the odd channel 
as seen from \fig{qwa0-d}. 
The main differences are that its intensity is overall suppressed 
and it has a larger energy scale (characterized by $\omega_{\vQ}$, 
see also \fig{d-wQ}). 
These features of the even channel were actually reported in neutron 
scatting experiments.\cite{bourges97,dai99,fong00,pailhes03,pailhes04,stock05,pailhes06}  
The experiments, however, indicate a much stronger suppression 
of the even channel than that obtained in our calculation, especially 
in the low $\omega$ region. 
In fact, Fong {\it et al.}\cite{fong00} concluded that the even 
channel was fully gapped even at $T=200$K. 
Although the even channel could be suppressed more by taking a 
larger value of $J_{z}$ in \eq{RPAJ}, 
such a ``gap'' feature is not captured by the present theory. 

While the present theory captures prominent features of magnetic
excitations in YBCO, the line shapes of Im$\chi(\vQ,\,\omega)$ in 
\fig{wa0-d} are different from experimental 
data.\cite{bourges95,bourges98,fong00}  
The experimentally observed peak structure, which is interpreted 
as resonance, is much broader. 
From the experimental viewpoint it is not evident that the 
resonance is really situated below the continuum,
although we have interpreted it as an ingap collective mode 
especially for the odd channel (\fig{d-wQ}). 
Several experimentalists \cite{bourges95,fong00,dai01} extract the
gap from the low-energy tail of the resonance, such that the gap 
becomes smaller than the resonance energy by definition. 
From the present theoretical viewpoint, however, the gap should 
be defined from the lower edge of the continuum.

\subsection{Anisotropy of magnetic excitations} 

Hinkov {\it et al.}\cite{hinkov04} presented a clear $\vq$ map 
of magnetic spectral intensity for untwinned YBa$_{2}$Cu$_{3}$O$_{6.85}$, 
which showed that the peak intensity at $\vq=(\pi\pm 2\pi\eta_{x},\,\pi)$ 
is larger than that at $\vq=(\pi,\,\pi\pm 2\pi\eta_{y}) \,$. 
This map was obtained by fitting the experimental data to putative 
$\delta$-peaks in $\vq$ space smeared by a spectrometer resolution 
function.\cite{hinkov05}   
In the present model (see \fig{qa5-d}), the peak intensity 
difference for $\omega <\omega_{\vQ}^{\rm res}$ is less sizable 
below $\D\sim0.15$ than the difference extracted from the fit
to the experimental data. 
Hinkov {\it et al.}\cite{hinkov04} also reported that the $\vq$-integrated 
spectral weight difference between along 
$\vq=(\pi\pm 2\pi\eta_{x},\,\pi)$ and $\vq=(\pi,\,\pi\pm 2\pi\eta_{y})$  
became  apparently larger with decreasing $\omega$ below the 
resonance energy. 
This feature is not manifest in our results shown in 
Figs.~\ref{qa5-w}(c)-(e).  

These discrepancies may result from details of the band structure. 
LDA calculations suggest\cite{andersen05} that seemingly small effects, 
such as long-range hopping integrals beyond third 
neighbors and the buckling of oxygen atoms, 
may affect the dispersion near the Fermi level without
necessarily leading to strong shifts of the Fermi surface. 
The recently observed kink structure of the band dispersion near
the Fermi level\cite{damascelli03} should also be taken into
account. 
The influence of band structure features on the IC peak intensity 
difference follows also from theoretical studies. 
Zhou and Li\cite{zhou0405} showed that the direct coupling to the chain 
band contributes strongly to the peak intensity difference. 
Two recent phenomenological calculations with different band parameters 
also revealed the sensitivity to band structure details: 
in one case 22\% band anisotropy\cite{kao05} was required to account for 
the peak intensity difference reported by Hinkov {\it et al.}\cite{hinkov04}, 
in the other only 6\%.\cite{eremin05}  
There may also be ambiguities in the analysis of the experimental data, 
since the anisotropy of the IC peaks was deduced after a 
background correction, and the larger anisotropy was reported when the 
background became dominant, that is when the magnetic signals became 
very weak. 
If there is a magnetic signal hidden in the background, one might
overestimate the IC peak intensity difference. 

\begin{figure}
\centerline{\includegraphics[width=0.4\textwidth]{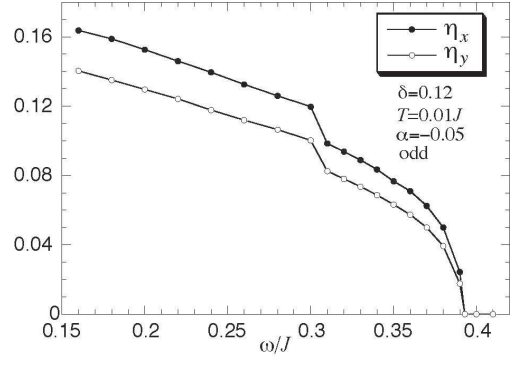}}
\caption{Energy dependence of $\eta_{x}$ and $\eta_{y}$ for 
$\D=0.12$, $T=0.01J$, and $\alpha=-0.05$ in the odd channel. 
The jump of $\eta_{x}$ and $\eta_{y}$ at $\omega\approx 0.3J$ is due to 
fine peak structures of Im$\chi(\vq,\,\omega)$.  
}
\label{eta-w}
\end{figure}

Since the anisotropy of the IC peak intensity is sensitive to details 
of the model and to ambiguities in the experimental data analysis, 
it seems advantageous to consider the anisotropy of the incommensurability, 
$\Delta\eta = \eta_{x}-\eta_{y}$, as a more robust measure of the 
anisotropy of Im$\chi(\vq,\,\omega)$ at low $T$ for 
$\omega < \omega_{\vQ}^{\rm res}$.  
We have consistently obtained $\Delta\eta \gtrless 0$ for 
$\alpha \lessgtr 0$ for all studied parameter sets, and moreover 
$\Delta\eta$ turns out to be proportional to $\alpha$ 
in good approximation at least up to $|\alpha|\sim 0.1$ without 
appreciable $\omega$ dependence except near $\omega_{\vQ}^{\rm res}$ 
(\fig{eta-w}).  
Our result, $\Delta\eta \approx 0.02-0.03$ for $\alpha=-0.05$ 
(\fig{eta-w}), is comparable with experimental data for YBa$_{2}$Cu$_{3}$O$_{6.6}$ 
and not inconsistent with the data for YBa$_{2}$Cu$_{3}$O$_{6.85}$ 
also.\cite{hinkov04,hinkov05} 

The negative sign of $\alpha$ corresponds to $t_{x}<t_{y}$, that is,
the opposite of what one would expect from the in-plane lattice 
constant difference. 
In agreement with LDA band calculations,\cite{andersen95} this implies
that the chains are crucial to the in-plane anisotropy in YBCO and thus 
to the understanding of the anisotropy of Im$\chi(\vq,\,\omega)$. 
The importance of chain effects was already discussed by Zhou and 
Li\cite{zhou0405} in a different context. 
They analyzed direct coupling between CuO chains and CuO$_{2}$ planes. 
Such a direct coupling, 
however, should be small enough to ensure that Im$\chi(\vq,\,\omega)$ 
is well characterized by a bilayer model to be consistent with 
the observed $q_{z}$ modulation.\cite{tranquada92,mook93,fong9596,
dai96,bourges96,bourges98,pailhes03,stock04,pailhes06}  
We have therefore interpreted chain effects as mainly renormalizing
the in-plane band anisotropy. 
We note that an LDA calculation\cite{andersen95} predicts that chain 
effects are not important to the anisotropy of $t$ in the double 
chain compound YBa$_{2}$Cu$_{4}$O$_{8}$, for which we thus expect 
$t_{x}>t_{y}$. 
Our theory then predicts $\Delta\eta = \eta_{x}-\eta_{y} <0$ 
for $\omega < \omega_{\vQ}^{\rm res}$ at low $T$. 

For $\omega > \omega_{\vQ}^{\rm res}$ we have found a pronounced 
spectral weight anisotropy, which however strongly depends on 
$\omega$ [Figs.~\ref{qa5-w}(f)-(h)]. 
Stock {\it et al.}\cite{stock05} performed high energy neutron 
scattering experiments for partially untwinned YBa$_{2}$Cu$_{3}$O$_{6.5}$ and 
obtained $\vq$ maps of the spectral weight. 
They reported nearly isotropic spectral weight distribution. 
However, their data were integrated over intervals $\pm 7.5$meV 
along the energy axis, which may smear out the predicted anisotropy 
in Figs.~\ref{qa5-w}(f)-(h). 
In addition, anisotropies may be underestimated if the untwinning 
of the crystal is not complete.   
On the other hand, Hinkov {\it et al.}\cite{hinkov06} observed an 
anisotropy of IC peaks at several choices of energy for  
$\omega> \omega_{\vQ}^{\rm res}$ in almost fully untwinned 
YBa$_{2}$Cu$_{3}$O$_{6.6}$, while full $\vq$ 
maps have not yet been obtained.  
Further neutron scattering data for $\omega > \omega_{\vQ}^{\rm res}$
would be useful. 

At relatively high $T$ 
we have found an 
enhanced anisotropy of Im$\chi(\vq,\,\omega)$ [Figs.~\ref{qa5-T}(c) 
and (d)]. This anisotropy is characterized by the difference of the  
broad commensurate peak width between the $q_{x}$ direction and 
the $q_{y}$ direction. This anisotropic peak width was actually 
observed in the most recent experiment for YBa$_{2}$Cu$_{3}$O$_{6.6}$ in the 
pseudogap regime.\cite{hinkov06}

\subsection{Relevance of ${\boldsymbol d}$FSD correlations}

A sizable anisotropy of Im$\chi(\vq,\,\omega)$ was reported  
for underdoped materials, YBa$_{2}$Cu$_{3}$O$_{6.5}$\cite{stock04} and 
YBa$_{2}$Cu$_{3}$O$_{6.6}$.\cite{hinkov04,hinkov06} 
To understand this behavior, the bare band anisotropy, without 
an enhancement due to $d$FSD correlations, is not sufficient. 
The magnetic anisotropy due to the bare band structure anisotropy
decreases at lower $\D$ for fixed $\alpha$, as seen in the right 
panels of \fig{qa5-d}. 
One cannot expect a larger $\alpha$ at lower $\D$ since the 
crystal structure changes from orthorhombic to tetragonal for 
$y\lesssim 6.4$, and is accompanied by the disappearance of
the CuO chains.\cite{jorgensen90} 
On the other hand, the presence of $d$FSD correlations provides 
a natural explanation for the observed anisotropy at low $\D$; 
the correlation effects can yield an enhanced anisotropy of  
magnetic excitations at low $\D$ (\fig{qa5-d}). 
While we have treated $d$FSD correlations in the slave-boson 
mean-field approximation to the $t$-$J$ model, they were shown 
to be present also in a recent exact diagonalization 
study.\cite{miyanaga06} 
The anisotropy of Im$\chi(\vq,\,\omega)$ in optimally doped 
YBa$_{2}$Cu$_{3}$O$_{6.85}$\cite{hinkov04}, on the other hand, 
can be understood qualitatively 
already from the bare band anisotropy [Figs.~\ref{qa5-d}(c') 
and (d')], which is however further enhanced by $d$FSD 
correlations as seen in Figs.~\ref{qa5-d}(c) and (d). 

$d$FSD correlations also provide a natural scenario to account 
for the observed enhanced anisotropy of Im$\chi(\vq,\,\omega)$ 
in the pseudogap phase of underdoped YBa$_{2}$Cu$_{3}$O$_{6.6}$.\cite{hinkov06} 
As seen in Figs.~\ref{qa5-T} (c), (c'), (d), and (d'), the 
anisotropy of Im$\chi(\vq,\,\omega)$ is substantially enhanced 
compared with the bare anisotropy effect especially at 
relatively high temperatures. 
We note that the obtained anisotropy at high $T$ in the left 
panel of \fig{qa5-T} may be underestimated in the present 
calculation since we have assumed an isotropic hopping amplitude
of bosons $\bra b_{\vr}^{\dagger}b_{\vr'}\ket = \D$. 
Above the superconducting transition temperature, 
the bosons are not really condensed at 
the bottom of the band and $\bra b_{\vr}^{\dagger}b_{\vr'}\ket$ 
can become anisotropic, contributing thus to an enhancement of 
the anisotropy of magnetic excitations.

Due to $d$FSD correlations, we can expect a relatively large 
anisotropy at low $\D$ and high $T$ in orthorhombic YBCO. 
A large anisotropy was actually reported in resistivity 
measurements a few yeas ago.\cite{ando02}   
The data, however, were interpreted in terms of partial
spin-charge stripe order, often referred to as (electronic) 
nematic order.\cite{kivelson98}  
Although the nematic order has the same symmetry as the $d$FSD, 
namely orientational symmetry breaking of the square lattice,  
the underlying mechanism is different. 
$d$FSD correlations come from forward scattering processes of 
quasiparticles close to the FS near 
$(\pi,\,0)$ and $(0,\,\pi)$,\cite{yamase00,metzner00} 
while a spin-charge stripe requires correlations with a
large momentum transfer such as antiferromagnetism. 
Our scenario offers a different route to understand the 
anisotropy of the resistivity\cite{ando02} as well as the 
anisotropy of the magnetic excitation 
spectrum\cite{stock04,hinkov04,hinkov06} on the basis of 
a microscopic model calculation.

\subsection{Fermi surface and superconducting gap anisotropy} 

Because of $d$FSD correlations, the FS of the antibonding band 
can easily open in presence of a small bare band anisotropy 
as shown in \fig{FS}. 
The Fermi surface topology depends in particular on the values of 
$\alpha$ and $t_{z}/t$. 
Further efforts to determine the FS shape in YBCO\cite{schabel98,lu01} 
will serve to extract these parameters, which then gives insights on 
chain band effects as well as on the present scenario for the 
anisotropy of magnetic excitations. 
As shown in \fig{density}, the hole density of the antibonding band 
is significantly larger than that of the the bonding band. 
This self-consistent result may be useful for mapping the FS in 
future ARPES studies.  

Although the inner FS can change its topology with $\alpha$, 
such a topology change does not strongly affect the anisotropy of 
Im$\chi(\vq,\,\omega)$. We have not observed an enhanced $\Delta\eta$ 
when the inner FS opens, but a simple linear relation 
$\Delta\eta \propto \alpha$ in good approximation at low $T$ for 
$\omega < \omega_{\vQ}^{\rm res}$. 

The present self-consistent calculation yields 
$\Delta_{x}>|\Delta_{y}|$ for $\alpha<0$, that is, $\Delta_{s} > 0$ 
[\fig{meana5}(a)], which partially compensates the bare anisotropy 
of $J_{\vr\,\vr'}$ [see \eq{insc}]; 
note that $\Delta_{\tau}$ is the RVB pairing amplitude and 
the true gap magnitude is given by $J_{\tau}\Delta_{\tau}$. 
Hence we obtain $J_{x}\Delta_{x} < |J_{y}\Delta_{y}|$  
with an anisotropy a bit smaller than $|2\alpha|$, about
$7\%$ for $\D=0.10-0.20$ and $\alpha=-0.05$. 
The relation $J_{x}\Delta_{x}<|J_{y}\Delta_{y}|$ was actually 
reported in a Raman scattering experiment for the optimally doped 
and overdoped YBCO.\cite{limonov00} 
Note that this gap anisotropy does not lead to a gap anisotropy 
between $\vk=(\pi,\,0)$ and $(0,\,\pi)$, 
which comes from $\Delta_{z}$ in the present theory,
as seen from Eqs.~(\ref{insc}) and (\ref{outsc}).  
While the magnitude of $\Delta_{z}$ might look sizable in 
\fig{meana5} (c) especially for high $\D$, $\Delta_{z}$ is 
multiplied by $J_{z}$ in \eq{outsc}. 
The resulting anisotropy between $\Delta_{\vk=(\pi,\,0)}$ and 
$\Delta_{\vk=(0,\,\pi)}$  becomes very small, less than $1\%$ 
for our parameters.
Recent ARPES data\cite{lu01} reported an anisotropy of about 
$50\%$, which cannot be understood in the present model.

\section{Conclusion}

We have performed a comprehensive analysis of the bilayer 
$t$-$J$ model in the slave-boson mean-field scheme. 
After determining the mean fields self-consistently, we have 
calculated the dynamical magnetic susceptibility in a renormalized 
RPA.   
We have shown that prominent features of magnetic excitations of 
YBCO are captured by this scheme: 
(i) the IC and DIC signals only for $T<T_{\rm RVB}$, 
(ii) the collective mode for $T<T_{\rm RVB}$,
(iii) its downward dispersion for $\omega \lesssim \omega_{\vQ}^{\rm res}$, 
(iv) the $\D$ dependence of $\omega_{\vQ}^{\rm res}$,
(v) the overdamped collective mode for 
$\omega\gtrsim \omega_{\vQ}^{\rm res}$ with an upward dispersion, 
(vi) robustness of this high energy dispersive feature to $T$, and 
(vii) spectral weight suppression of the even channel.

In particular, the present theory, which includes $d$FSD 
correlation effects in a self-consistent manner, 
provides a natural scenario to understand the observed anisotropy of 
Im$\chi(\vq,\,\omega)$: 
(i) appreciable anisotropy also for low $\D$,  
and (ii) enhanced anisotropy  at relatively high $T$ $(\sim T_{\rm RVB})$. 

Although magnetic excitations of high-$T_{c}$ cuprates are 
often discussed in terms of spin-charge stripes after the proposal 
by Tranquada {\it et al.},\cite{tranquada95}   
the present comprehensive study indicates that conventional particle-hole 
scattering processes around the FS are essential to magnetic excitations. 
An additional insight from this study is the importance of $d$FSD 
correlations for the anisotropy of Im$\chi(\vq,\,\omega)$. 
$d$FSD correlations are due to forward scattering interactions,
which were so far ignored in most theories.

\begin{acknowledgments}
We are grateful to O.K. Andersen, V. Hinkov, B. Keimer, D. Manske, 
and R. Zeyher for helpful discussions and for sharing some 
unpublished results with us. 
\end{acknowledgments}


\bibliography{main.bib}

\newpage

\clearpage

\end{document}